\documentclass{article}
\usepackage{PRIMEarxiv}

\usepackage[utf8]{inputenc} % allow utf-8 input
\usepackage[T1]{fontenc}    % use 8-bit T1 fonts
\providecommand{\bmsection}{\section}
\usepackage{url}            % simple URL typesetting
\usepackage{booktabs}       % professional-quality tables
\usepackage{amsfonts}       % blackboard math symbols
\usepackage{nicefrac}       % compact symbols for 1/2, etc.
\usepackage{microtype}      % microtypography
\usepackage{lipsum}
\usepackage{mathtools}
\usepackage{makecell}
\usepackage{stmaryrd}
\usepackage[numbers,sort&compress]{natbib}
\usepackage[dvipsnames]{xcolor} % in the preamble
\usepackage{float}
\usepackage{fancyhdr}       % header
\usepackage{graphicx}       % graphics
\usepackage{subcaption}
\graphicspath{{media/}}     % organize your images and other figures under media/ folder
\usepackage{siunitx}
\usepackage{listings}
\usepackage{lineno}
\usepackage{dsfont}
\usepackage[colorlinks = True, linkcolor = violet, citecolor = blue]{hyperref}       % hyperlinks
\usepackage[colorinlistoftodos,textsize=footnotesize]{todonotes}
\usepackage{pifont}

\definecolor{codegreen}{rgb}{0,0.6,0}
\definecolor{codegray}{rgb}{0.5,0.5,0.5}
\definecolor{codepurple}{rgb}{0.58,0,0.82}
\definecolor{backcolour}{rgb}{0.95,0.95,0.92}

\newcommand{\xmark}{\ding{55}}%

\lstdefinestyle{mystyle}{
    backgroundcolor=\color{backcolour},   
    commentstyle=\color{codegreen},
    keywordstyle=\color{magenta},
    numberstyle=\tiny\color{codegray},
    stringstyle=\color{codepurple},
    basicstyle=\ttfamily\footnotesize,
    breakatwhitespace=false,
    xleftmargin=1em,
    xrightmargin=1em,
    breaklines=true,                 
    captionpos=b,                    
    keepspaces=true,                 
    numbers=left,                    
    numbersep=5pt,                  
    showspaces=false,                
    showstringspaces=false,
    showtabs=false,                  
    tabsize=2
}

\graphicspath{{./figures_article/}}

\newcommand{\bbeta}{\boldsymbol{\beta}}
\newcommand{\btheta}{\boldsymbol{\theta}}

\newcommand{\bmu}{\boldsymbol{\mu}}
\newcommand{\beeta}{\boldsymbol{\eta}}
\newcommand{\bOmega}{\boldsymbol{\Omega}}
\newcommand{\y}{\mathbf{y}}

\newcommand{\X}{\mathbf{X}}
\newcommand{\I}{\mathbf{I}}

\newcommand{\N}{\mathcal{N}}
\newcommand{\D}{\mathcal{D}}

\title{Joint Bayesian Inference of Genetic Effect Sizes and PK Parameters in Nonlinear Mixed-Effects Models}

\author{Julien Martinelli\\ELLIS Institute Finland\\Aalto University\\ Espoo, Finland\\\texttt{julien.martinelli@aalto.fi}
\And Ibtissem Rebai\\Certara\\ Paris, France\\ \texttt{ibtissem.rebai@certara.com}
\And David W. Haas\\ Vanderbilt University Medical Center\\Nashville, Tennessee, USA \\ \texttt{david.haas@vumc.org}
\And Julie Bertrand\\UMR1137 IAME\\Université Paris Cité, Université Paris Sorbonne Nord, INSERM\\ Paris, France\\\texttt{julie.bertrand@inserm.fr}}

\begin{document}
%\linenumbers

\maketitle

\begin{abstract}
High-dimensional genetic covariate selection in population pharmacokinetic (PK) models is challenging due to the cohort's restricted size and high correlation among single-nucleotide polymorphisms (SNPs).
We propose a fully Bayesian, single-stage framework that jointly infers nonlinear mixed effect model (NLMEM) parameters and SNP effect sizes, providing coherent posterior uncertainty and inclusion summaries within a single model fit.
We compare five sparsity-inducing priors---Spike-and-Slab, Hierarchical Lasso, Regularized Horseshoe, R2--D2, and the $\ell_1$-ball---calibrated through effect-size and sparsity targets.
In simulations, all priors showed low false-discovery rates around $0$--$0.08$ under the null, and recovered the causal signal under the alternative, with peak $F_1$ scores around $0.8$--$0.85$ under reasonable inclusion cutoffs.
Spike-and-Slab was especially attractive because it provides analytical posterior inclusion probabilities directly, while among priors requiring tolerance-based proxy inclusion summaries, the $\ell_1$-ball combined similarly strong recovery with the most stable behavior across tolerance values.
On genetic and PK data from the ANRS 12154 study in 129 Cambodians living with HIV and receiving nevirapine, posterior predictive checks indicated adequate calibration and PK parameter inference remained stable across priors.
While the dominant signal was robust across priors, additional candidate SNPs showed only partial agreement in ranking and more prior-sensitive effect-size estimates.
These results support Bayesian variable selection within joint NLMEM as a principled approach for pharmacogenetic analyses when uncertainty quantification and regularization are central.
\end{abstract}

\section{Introduction}

Inter-individual variability in drug exposure and response is a central concern in pharmacology and drug development.
Patients receiving the same dose may exhibit markedly different concentration--time profiles and clinical outcomes, reflecting heterogeneity in physiological, environmental, and genetic factors.
Pharmacogenetics (PGx) seeks to explain part of this variability by identifying genetic variants associated with pharmacokinetic (PK) or pharmacodynamic (PD) parameters, thereby improving mechanistic understanding of variability and informing individualized treatment strategies~\cite{relling2015pharmacogenomics}.

%In drug development and clinical pharmacology, PGx effects are naturally investigated through their impact on PK/PD model parameters.
Nonlinear mixed-effects (NLME) models provide a principled framework to quantify population-level parameters, between-subject variability, and residual noise from PK/PD data while accommodating sparse and unbalanced designs.
Embedding genetic covariates directly within an NLME model provides a principled alternative to two-stage procedures (e.g., association testing on empirical Bayes estimates), and has been shown to improve the calibration and power to detect pharmacogenetic covariates compared to approaches relying on observed concentrations or metrics derived from non-compartmental analyses. \citep{tessier2015comparison}.

Yet, PGx covariate selection is often high-dimensional: many variants are evaluated in cohorts that are modest in size relative to the number of candidate SNPs.
In practice, pharmacometric workflows have long relied on stepwise covariate model building and related frequentist strategies, which can be unstable and subject to selection-induced bias; uncertainty quantification can also become hard to interpret after selection. \citep{hutmacher2015covreview}
Recent work has therefore explored penalized and stability-enhanced methods for NLME covariate selection \citep{naveau2024bayesian,gabaut2025stability,caillebotte2025estimationvariableselectionhigh}, but principled post-selection inference remains delicate in general \citep{pnasselectiveinference}, and standard error estimation in NLME settings can be challenging under sparse designs or small cohorts. \citep{broeker2020assessing,bertrand2012some}

Bayesian inference offers an appealing route to address these limitations by treating covariate effects, random effects, and observation noise jointly, and by propagating uncertainty coherently to downstream quantities and predictions.
Bayesian population PK/PD modeling has a long history (e.g., via WinBUGS/PKBugs~\cite{lunn2002pkbugs}), but its routine use in NLME settings was historically constrained by the computational cost of sampling-based inference.
This barrier has been progressively lowered by modern gradient-based algorithms---Hamiltonian Monte Carlo (HMC) and the No-U-Turn Sampler (NUTS)~\citep{neal2011mcmc,nuts}---which exploit model geometry to explore high-dimensional posteriors efficiently, and by differentiable probabilistic programming frameworks that automate gradient computation and enable scalable implementations of HMC/NUTS~\citep{bingham2019pyro,phan2019composable,jax2018github}.
A key benefit of the Bayesian paradigm is the ability to incorporate mechanistic or empirical prior information so that additional computation yields tangible gains in regularization, identifiability, and interpretability, as illustrated in dose-finding contexts, nonlinear joint models, and finite-sample uncertainty analyses. \citep{ursino2017dosefinding,kerioui2020bayesian,guhl2024uncertainty}

In this work, we study Bayesian variable selection for pharmacogenetic covariates in NLME models and compare five complementary families of sparsity-inducing priors.
Rather than committing to a single selection mechanism, we consider priors that span distinct design principles: mixture-based selection with analytical posterior inclusion probabilities (spike-and-slab),
convex projection yielding exact zeros without discrete indicators ($\ell_1$-ball), 
global--local shrinkage (hierarchical lasso and regularized horseshoe), and variance-explained parametrization (R2-D2)~\citep{spike-slab,l1ball,peltola_2014_horseshoe,Piironen_2017,r2d2}.
Each prior strongly concentrates most SNP effects near zero while retaining enough tail mass to accommodate occasional non-negligible effects---a key requirement for pharmacogenetic architectures dominated by a small number of variants.
Recent pharmacometric work has also highlighted the practical appeal of continuous shrinkage priors, such as the regularized horseshoe, as a convenient alternative to stepwise covariate modeling. \citep{pourzanjani2026rhs}

%The remainder of the paper is organized as follows.
Hereby, we first introduce the NLME model and the PGx parametrization used throughout, together with a prior calibration strategy reflecting expected effect sizes in PGx settings.
We then present the five sparsity priors and the computational approach used for posterior inference.
Next, we conduct a simulation study to assess calibration, selection accuracy, and robustness across priors and genetic architectures.
Finally, we apply the framework to the 12154 ANRS study of nevirapine PGx in Cambodians living with Human immunodeficiency virus (HIV) type 1, where a well-established variant is modeled explicitly and the remaining SNP effects represent additional genetic contributions beyond this dominant determinant.
We conclude with practical guidance on prior choice, diagnostics, and limitations.

\section{Methods}

\subsection{Model and notations}\label{sec:model}

%Throughout, bold lowercase symbols (e.g.\ $\mathbf{x}$) denote vectors and bold uppercase symbols (e.g.\ $\mathbf{X}$) denote matrices.

Let $\mathbf{y}_i$ denote the vector of individual $i=1,\dots,N$ observed concentrations at vector time $\mathbf{t}_i$, such that $(\mathbf{t}_i,\mathbf{y}_i)=\{(t_{il},y_{il})\}_{l=1}^{n_i}$, with $n_i$ the number of sampling times $t_{il}$ of patient $i$ which we describe with the statistical model below:
\begin{align}
y_{il} &= f\!\big(\boldsymbol{\phi}_i; t_
{il}, D\big) + v_{il}, \\
\boldsymbol{\phi}_i &= \exp(\boldsymbol{\mu} + \boldsymbol{\eta}_i),
\end{align}
where $f(\cdot)$ is a mathematical function, corresponding to a PK compartmental model, nonlinear in its parameters $\boldsymbol{\phi}_i$ notably composed of $\beeta_i \sim \mathcal{N}(\mathbf{0},\bOmega)$ the individual random effects and with $\mathbf{v}_{i} \sim \mathcal{N}(0, \sigma^2 \mathbf{I}_{n_i})$ the residual errors. Population-level parameters are $\btheta=(\bmu,\bOmega,\sigma)$: the fixed-effects vector $\bmu$, interindividual covariance $\bOmega$, and residual variance $\sigma^2$.

Here, we consider a genetic effect on the $h^{\text{th}}$ element of the vector $\phi_i$.
All $p$ single-nucleotide polymorphisms (SNPs) are coded in $\{0,1,2\}$ according to the number of minor alleles and standardized to have mean zero and unit variance.
Let $\mathbf{X} \in \mathbb{R}^{N \times p}$ denote the resulting genotype matrix, with entry $X_{i,j}$ for subject $i$ and SNP $j$, and let $\boldsymbol{\beta} = (\beta_1,\ldots,\beta_p)^\top \in \mathbb{R}^p$ denote the corresponding vector of SNP effects.
Genetic effects are incorporated on the log-scale of the $h^{\text{th}}$ PK parameter, so that for subject $i$
\begin{equation}
\log \phi_{i,h} = \hat{\mu}_h + \sum_{j=1}^{p} X_{i,j} \beta_j + \eta_{ih}.
\label{eq:lrcl}
\end{equation}

Posterior inference is performed using NUTS, an adaptive variant of Hamiltonian Monte Carlo~\citep{neal2011mcmc, nuts}. 
The latter augments parameters $q$ with momenta $m$ and simulates Hamiltonian dynamics for the energy
\begin{equation}
H(\mathbf{q},\mathbf{m}) = -\log \pi(\mathbf{q}) + \frac{1}{2}\mathbf{m}^\top \mathbf{M}^{-1}\mathbf{m},
\end{equation}
using a symplectic leapfrog integrator with a specific step size $\epsilon$ and a mass matrix $\mathbf{M}$, followed by a Metropolis correction. This yields long-distance, gradient-informed proposals that avoid the random-walk behavior of Gibbs/Metropolis methods and mix efficiently in high dimensions. NUTS removes the need to hand-tune the trajectory length by building a binary tree of leapfrog steps and stopping automatically when the simulated path makes a \emph{U-turn}~\citep{nuts}.

\subsection{Bayesian variable selection approaches}

To infer which SNPs contribute to the variability in the PK parameter of interest, we consider various sparsity-inducing priors on $\bbeta$. 
To make hyperparameters comparable across experiments and across priors, we factor out the natural scale of genetic effects. 
Let $\omega_{h}$ denote the interindividual standard deviation on the log-scale of the $h^{\mathrm{th}}$ element of $\boldsymbol{\phi}_i$ (Eq.~\ref{eq:randomeffects}). 
Conceptually, we may write $\beta_j = \omega_{h}\,\tilde\beta_j$, where $\tilde\beta_j$ is dimensionless. 
In practice, we keep $\beta_j$ as the model parameter and parameterize each prior so that 
$\mathrm{Var}(\beta_j \mid \cdot)$ is proportional to $\omega_{h}^2$, 
which is equivalent to placing sparsity-inducing priors on the rescaled coefficients $\tilde\beta_j$. 
Under this parameterization, prior calibration depends on $(N,p)$ and on the targeted sparsity or effect size, 
but is invariant to the overall scale of the phenotype. For all priors below, $j \in \{1,\dots,p\}$ indexes SNPs.

\paragraph{Spike-and-Slab~\citep{spike-slab}.}

This prior can be expressed as a hierarchical model~\citep{Piironen_2017}:
\begin{equation}
\label{eq:sas}
\begin{aligned}
&\beta_j \mid z_j 
\sim 
\begin{cases}
\mathcal{N}\!\left(0,\ \omega_h^2 \sigma_{\text{spike}}^2\right), 
& z_j = 0 \quad \text{(spike)}, \\[4pt]
\mathcal{N}\!\left(0,\ \omega_h^2 \sigma_{\text{slab}}^2\right), 
& z_j = 1 \quad \text{(slab)},
\end{cases} \\
&\text{with inclusion probability } \quad 
z_j \mid \gamma \sim \mathrm{Bernoulli}(\gamma),\\
&\text{and global sparsity parameter } \quad 
\gamma \sim \mathrm{Beta}(\gamma_a,\gamma_b).
\end{aligned}
\end{equation}
%
% Equivalently, $\beta_j \mid \theta \sim (1-\theta)\mathcal{N}(0,\sigma_{\text{spike}}^2)+\theta\mathcal{N}(0,\sigma_{\text{slab}}^2)$.
% This produces exact zeros only if you choose a point-mass spike; here it’s a very narrow normal “near-zero” component.

Here, $\gamma \in (0,1)$ denotes the global inclusion probability and 
$0 < \sigma_{\text{spike}} \ll \sigma_{\text{slab}}$.
Each coefficient $\beta_j$ is governed by a latent inclusion variable $z_j$,
which selects between a strongly concentrated “spike” component and a more diffuse “slab” component.
The global parameter $\gamma$ controls prior sparsity, with expected prior inclusion probability
\[
\mathbb{E}[\gamma] = \frac{\gamma_a}{\gamma_a + \gamma_b}.
\]
Because the spike is a narrow Gaussian distribution rather than a point mass at zero,
the prior strongly shrinks small effects toward zero but does not enforce exact zeros.

The $z_j$ are discrete, whereas classical state-of-the-art MCMC samplers operate on smooth densities. %(see Section~\ref{sec:hmc}). 
Rather than sampling the $z_j$'s, we integrate them out:
\begin{equation}
p(\beta_j\mid\gamma)
=\sum_{z_j\in\{0,1\}} p(\beta_j\mid z_j)p(z_j\mid\gamma)
=(1-\gamma)\N(\beta_j;0,\sigma_{\text{spike}}^2)+\gamma\N(\beta_j;0,\sigma_{\text{slab}}^2),
\end{equation}
so that NUTS only moves on the continuous variables $(\bbeta,\gamma)$. Moreover, marginalization reveals one of the key advantages of the spike-and-slab prior: access to the \emph{posterior inclusion probability} $p(z_j=1\mid \beta_j,\gamma)$ for each SNP in analytical form for each MCMC draw:
\begin{equation}
p(z_j=1 \mid \beta_j,\gamma)
=\frac{\gamma\N(\beta_j;0,\sigma_{\text{slab}}^2)}
{\gamma\N(\beta_j;0,\sigma_{\text{slab}}^2)+(1-\gamma)\N(\beta_j;0,\sigma_{\text{spike}}^2)}.
\label{eq:pipss}
\end{equation}

\paragraph{$\ell_1$-ball prior~\cite{l1ball}.}
In this prior, we have
\begin{equation}
\label{eq:l1ball}
\begin{aligned}
&\tilde{\boldsymbol{\beta}}
=
\operatorname*{argmin}_{\|\tilde{\boldsymbol{\beta}}\|_1 \le \tilde r}
\frac12
\|\tilde{\boldsymbol{\beta}} - \tilde{\boldsymbol{\xi}}\|_2^2, \\
\text{with latent vector } \quad 
&\tilde{\xi}_j \sim \mathrm{Laplace}(0,b_\xi),\\
\text{and budget parameter } \quad 
&\tilde r \sim \mathrm{Exp}(\lambda_r),
\end{aligned}
\end{equation}
and the regression coefficients are obtained via
\[
\boldsymbol{\beta} = \omega_h \tilde{\boldsymbol{\beta}}.
\]

Equivalently, the solution has the soft–thresholding form
\[
\tilde{\beta}_j
=
\mathrm{sign}(\tilde{\xi}_j)
\bigl(|\tilde{\xi}_j| - \tilde{\delta}\bigr)_+,
\]
where the common threshold $\tilde{\delta} \ge 0$ satisfies
\[
\sum_{j=1}^p (|\tilde{\xi}_j| - \tilde{\delta})_+
=
\begin{cases}
\|\tilde{\boldsymbol{\xi}}\|_1,
& \|\tilde{\boldsymbol{\xi}}\|_1 \le \tilde r, \\[6pt]
\tilde r,
& \|\tilde{\boldsymbol{\xi}}\|_1 > \tilde r.
\end{cases}
\]

If $\tilde{\boldsymbol{\xi}}$ lies inside the $\ell_1$-ball,
the projection leaves it unchanged.
Otherwise, all coordinates are uniformly shrunk by $\tilde{\delta}$,
and those with $|\tilde{\xi}_j| \le \tilde{\delta}$ become exactly zero.
Sparsity therefore arises through convex projection rather than mixture modeling.

The sparsity level is governed by the budget parameter $\tilde r$:
smaller values increase the threshold $\tilde{\delta}$ and induce more exact zeros.
Since $\tilde r \sim \mathrm{Exp}(\lambda_r)$, the expected active size satisfies
\[
\mathbb{E}[K]
\approx
\frac{\mathbb{E}[\tilde r]}{b_\xi}
=
\frac{1}{\lambda_r b_\xi}.
\]

\paragraph{Hierarchical Lasso~\citep{peltola_2014_horseshoe}.}
This prior can be expressed as a normal scale mixture: for $ j=1,\ldots,p$,
\begin{equation}
\begin{aligned}
\beta_j \mid \lambda_j,\tau &\sim \mathcal{N}\!\left(0,\ \omega_{h}^2\,\tau^2\lambda_j^2\right),\\
\text{with global scale } \quad \tau &\sim \mathcal{H}\N(\tau_0),\\
\text{and local scales } \quad \lambda_j &\sim \mathrm{Exp}(\rho).
\end{aligned}
\label{eq:betahl}
\end{equation}
with local factors $\lambda_j$ and a global shrinkage parameter $\tau$ drawn from an exponential distribution of parameter $\rho$
 and a Half-Normal distribution of parameter $\tau_0$, respectively.
 
\paragraph{Regularized Horseshoe~\citep{Piironen_2017}.}
%Let $\lambda_j$ be local scales, $\tau$ a global scale, and $c$ a slab scale. % With half-$t$ families (Cauchy when $\nu=1$):
%
In this prior we have
\begin{equation}
\begin{aligned}
    &\beta_j \sim \mathcal{N}\!\left(0,\ \tau^2 \tilde{\lambda}_j^{2}\,\omega_{h}^2\right)\\
    \text{with global scale } \quad &\tau \sim \text{half-}t(\nu_{\text{global}},\tau_0)\\
    \text{and slab scale } \quad &c \sim \text{half-}t(\nu_{\text{slab}},\sigma_{\text{slab}})\\
    \text{with local scales } \quad &\lambda_j \sim \text{half-}t(\nu_{\text{local}},1),\\
    \text{and } \quad &\tilde{\lambda}_j^2 = \frac{c^2\lambda_j^2}{c^2+\tau^2\lambda_j^2}.
\end{aligned}
\label{eq:betahs}
\end{equation}
In the regularized horseshoe prior $\tau$ and the $\lambda_j$ are drawn from a half-Student-$t$ distribution. This combination enforces strong shrinkage on most coefficients, while the heavy-tailed half-Student-$t$ allows a few effects to remain essentially unshrunk. 
%In this formulation, the prior variance is scaled by $\omega_{\text{h}}^2$, to anchor SNP effects to the variability of the PK parameter of interest.

% The regularized horseshoe prior induces sparsity through a global shrinkage parameter $\tau$ and local factors $\lambda_j$, each drawn from a half-Student-$t$ distribution. This combination enforces strong shrinkage on most coefficients, while the heavy-tailed half-Student-$t$ allows a few effects to remain essentially unshrunk.

\paragraph{R2-D2 prior~\citep{r2d2}.} In this prior we have
\begin{equation}
\begin{aligned}
    &\beta_j \sim \mathcal{N}\Bigl(0, \tau^2 \lambda_j~\omega_{\text{h}}^2\Bigr) \\
    \text{with global scale } \quad &\tau^2 = \frac{R^2}{1-R^2} \\
    \text{ where } \quad &R^2 \sim \mathrm{Beta}(a,b)\\
    \text{ and } \quad &(\lambda_1,\dots,\lambda_p) \sim \mathrm{Dirichlet}(\alpha/p,\ldots,\alpha/p)\\
\end{aligned}
\label{eq:betar2}
\end{equation}

Here, the variance of the $\beta_j$ reflects three factors: the global model fit through $R^2$, the overall variance of the genetic contribution to clearance $\omega_{\text{h}}$, similar to the horseshoe, and the local weights $\lambda_j$ which variance is distributed across predictors \emph{via} simplex weights. But most importantly, this prior is defined by a $\mathrm{Beta}(a,b)$ distribution on the coefficient of determination $R^2$.

\section{Simulation study}\label{sec:simstudy}
\subsection{Pharmacokinetic and genetic settings}

We simulated data using a classical one-compartment PK model with first-order absorption and elimination~\cite{weiss2023one}. After a dose $D$, the drug is absorbed in a homogeneous fashion with rate constant $k_a$ and eliminated with rate constant $k = CL/V$, where $CL$ is the clearance and $V$ is the compartment volume. 
For a dosing interval of 12 hours and sampling time $t$, the steady-state concentration is given by the Bateman function,
\begin{equation}
f(\boldsymbol{\phi};t,\text{D})
= \frac{D}{V}\frac{k_a}{k_a - k}
\left(
  \frac{e^{-k t}}{1 - e^{-12k}}
  -
  \frac{e^{-k_a t}}{1 - e^{-12k_a}}
\right),
\end{equation}
where $\boldsymbol{\phi_i}=(k_{a_i},CL_i,V_i)$ follow the mixed-effects construction on the log-scale described in section~\ref{sec:model}. Table~\ref{tab:simstudy} lists the fixed and random-effect scales used in all simulations.
Each dataset contained $N=400$ individuals receiving with a fixed dose $D=200$ and providing three sampling times at $t\in\{1,4,12\}$ hours.
For each individual, a vector of $p=134$ SNPs was drawn randomly from the observed SNP matrix from the 129 patients included in the ANRS 12154 study.
This preserves the cohort’s empirical minor-allele frequencies and linkage-disequilibrium structure, ensuring all polymorphisms are truly segregating in our population.
SNPs were encoded as allele counts and standardized \emph{columnwise} within each dataset.\\

We considered two hypotheses, $H_0$: no true genetic effect and $H_1$: one causal SNP impacting the drug clearance.
%\[
%X_{\cdot j}^{(d)} = \frac{G_{\cdot j}^{(d)}-\bar G_{\cdot j}^{(d)}}{\operatorname{sd}(G_{\cdot j}^{(d)})},
%\qquad j=1,\dots,p.
%\]
Under $H_1$, we selected SNP \texttt{rs3745274} as causal variant with $\beta_{rs3745274}=-0.13$ and $\beta_j=0$ for $j\neq rs_{3745274}$ while under $H_0$ we set $\bbeta\equiv 0$.
The SNP \texttt{rs3745274} and $\beta_{rs3745274}$ values were chosen to provide a good power for the detection of the SNP effect in the context of the simulations while remaining consistent with results found in the literature and more specifically in a previous analysis of the ANRS 12154 study~\cite{Bertrand2012}.

\begin{figure}[H]
    \centering
    \includegraphics[width=1\linewidth]{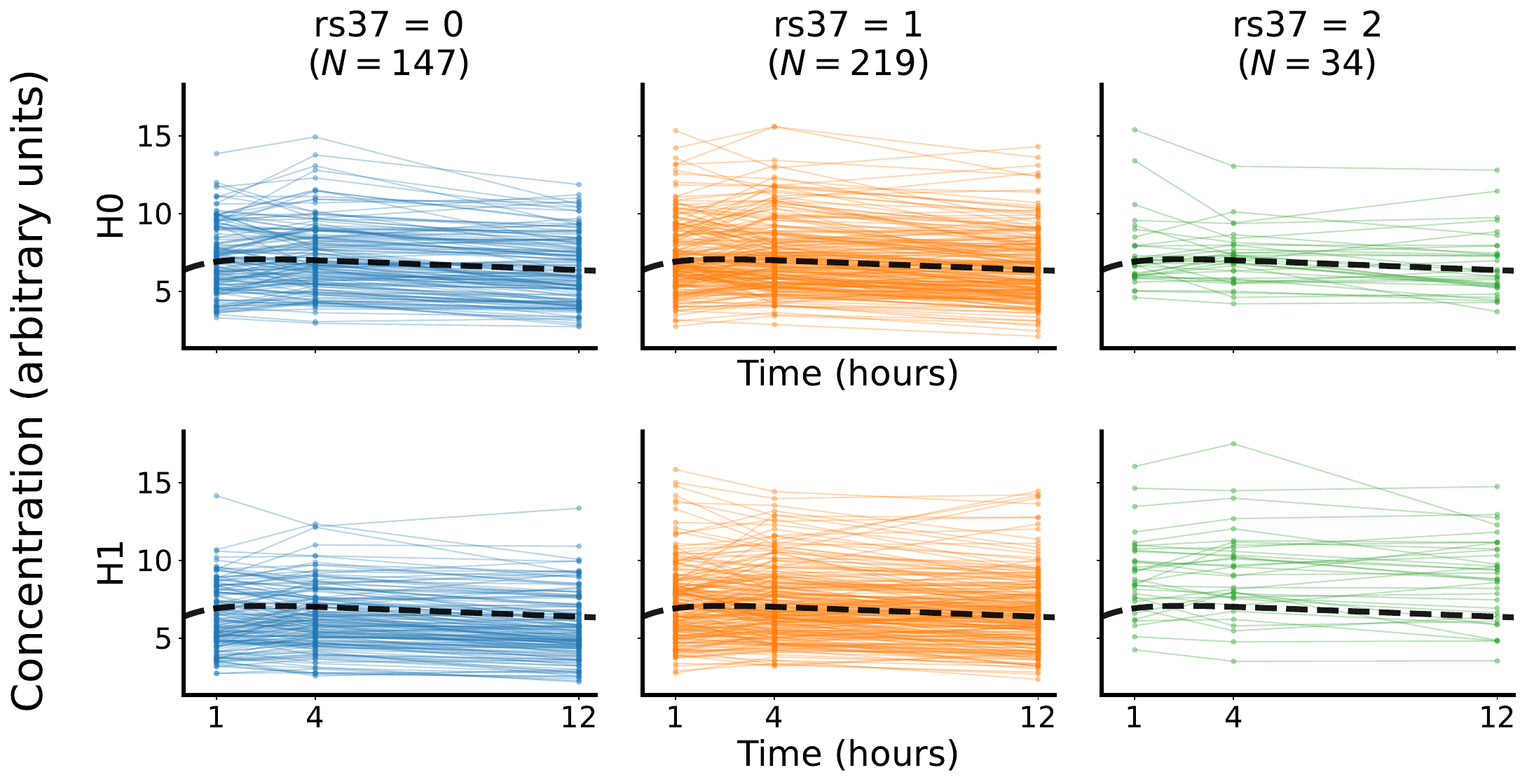}
    \caption{Observed temporal concentration profiles by genotype under $H_0$ (top) and $H_1$ (bottom). Each colored thin line shows one subject at the observed times (1, 4, 12h). The black dashed line is the latent population one-compartment model evaluated on a dense time grid.}
    \label{fig:fig2}
\end{figure}

Figure~\ref{fig:fig2} illustrates the simulated concentration profiles under $H_0$ (no genetic effects) and $H_1$ (one causal SNP, \texttt{rs3745274}). As expected, the profiles show substantial between-subject variability driven by the mixed-effects structure, and under $H_1$ the genetic signal induces only a moderate shift relative to this background variability, motivating the use of a model-based variable-selection procedure.

\begin{table}[H]
\centering
\caption{True parameter values used in the simulation study.}
\label{tab:true-params}
\begin{tabular}{l c @{\hspace{1.5em}} l c @{\hspace{1.5em}} l c}
\toprule
\multicolumn{2}{c}{Fixed effects ($\mu$, $\beta$)} &
\multicolumn{2}{c}{Inter-individual variability ($\omega$)} &
\multicolumn{2}{c}{Residual error ($\sigma$)} \\
\cmidrule(lr){1-2}
\cmidrule(lr){3-4}
\cmidrule(lr){5-6}
Parameter & Value & Parameter & Value & Parameter & Value \\
\midrule
$\exp(\mu_{CL})$ & $2.5$ 
& $\omega_{CL}$ & $0.3$
& $\sigma_1$ & $0.1$ \\

$\exp(\mu_{V_1})$ & $200$
& $\omega_{V_1}$ & $0.3$
&  &  \\

$\exp(\mu_{k_a})$ & $1.0$
& $\omega_{k_a}$ & $0.3$
&  &  \\

$\beta_{rs3745274}^{\star}$  [$H_1$] & $-0.13$
&  & 
&  &  \\

\bottomrule
\end{tabular}
\label{tab:simstudy}
\end{table}

\subsection{Hyperparameter tuning and prior calibration}

\paragraph{Priors over PK parameters}\label{sec:pkpriors}

To ensure positivity, PK parameters are modeled on the log-scale. 
Global parameters are assigned normal priors and interindividual variability terms are given independent half-normal priors,
% \begin{equation}
% \begin{aligned}
% \log \hat{k}_a &\sim \mathcal{N}(0.01,\sigma_{\text{PK}}^2),\\
% \log \hat{CL}  &\sim \mathcal{N}(0.9,\sigma_{\text{PK}}^2),\\
% \log \hat{V_1} &\sim \mathcal{N}(5.3,\sigma_{\text{PK}}^2),\\
% \log \sigma &\sim \mathcal{N}(\log 0.1,1)\\
% \omega_{k_a}, \omega_{CL}, \omega_{V_1} &\sim \mathcal{HN}(1.0).
% \end{aligned}
% \label{eq:interindivvar}
% \end{equation}
\begin{equation}
\begin{aligned}
\mu_{ka}   &\sim \mathcal{N}(0.01,\,\sigma_{\text{PK}}^2),\\
\mu_{CL}   &\sim \mathcal{N}(0.9,\,\sigma_{\text{PK}}^2),\\
\mu_{V_1}  &\sim \mathcal{N}(5.3,\,\sigma_{\text{PK}}^2),\\
\log \sigma &\sim \mathcal{N}(\log 0.1,\,1),\\
\omega_{ka},\omega_{CL},\omega_{V_1} &\sim \mathcal{HN}(1.0),
\end{aligned}
\label{eq:interindivvar}
\end{equation}
with \(\sigma_{\mathrm{PK}} = 0.25\).
Then, subject-specific random effects are drawn from a Gaussian
\begin{equation}
\begin{aligned}
u_{i,ka},u_{i,CL},u_{i,V_1} &\sim \mathcal{N}(0,1),\\
\eta_{i,ka} = \omega_{k_a}u_{i,ka},\quad
\eta_{i,CL} = \omega_{CL}&u_{i,CL},\quad
\eta_{i,V_1} = \omega_{V_1}u_{i,V_1}.
\end{aligned}
\label{eq:randomeffects}
\end{equation}

\paragraph{Priors over $\beta_j$.}
We encoded two pieces of prior knowledge:
(i) only a few SNPs are expected to be non-negligible
%(target $\tilde m \approx 5$ out of $p=134$)
, and
(ii) plausible genetic effects should correspond to realistic shifts in clearance on the natural scale.

Let $K$ denote the number of nonzero coefficients,
\begin{equation}
K := \sum_{j=1}^p \mathds{1}\{|\beta_j|\neq 0\}.
\end{equation}
We target $\mathbb{E}[K]\approx \tilde m$ with $\tilde m\approx 5$ out of $p=134$.

To calibrate effect magnitudes, recall that genetic effects enter the model on the log-clearance scale.
Given the inter-individual variability parameter $\omega_{CL}$, a coefficient of magnitude $|\beta|\approx 0.13$ corresponds to a moderate and clinically plausible multiplicative shift in clearance.
%Thus, the choice $|\beta|\approx 0.13$ is anchored to the variability scale of the pharmacokinetic parameter rather than selected arbitrarily.

For spike–and–slab and for the $\ell_1$–ball prior, $K$ is the natural sparsity measure induced by the prior.
For global–local shrinkage priors (Hierarchical Lasso, Regularized Horseshoe, R2–D2), coefficients are almost never exactly zero; we therefore employ the effective model size proxy introduced by~\citep{Piironen_2017}.
Magnitude calibration is performed by placing sufficient prior tail mass around the target effect size $|\beta|\approx 0.13$.

% Sparsity-enforcing priors are calibrated to encode two pieces of domain knowledge before seeing the data:
% (i) only a few SNPs are expected to have non-negligible effects (targeting $\tilde{m} \approx 5$ out of $p=134$),
% and (ii) a plausible effect size under $H_1$ is about $|\beta|\approx 0.13$.
% We quantify sparsity using the \emph{effective number of nonzero coefficients}~\citep[Equation 3.7]{Piironen_2017}
% \begin{equation}
% m_{\mathrm{eff}}(\bbeta) \;=\; \sum_{j=1}^p (1-\kappa_j), 
% \qquad \kappa_j\in(0,1) \text{ the shrinkage multiplier for } \beta_j,
% \label{eq:meff}
% \end{equation}
% which equals $\sum_j z_j$ for spike–and–slab (Equation~\ref{eq:sas}), and for continuous–shrinkage priors
% is defined via the prior shrinkage of each coefficient (see below).

\underline{Spike–and–slab prior.}

%This mixed-effects formulation captures both the typical pharmacokinetic behavior in the population, represented by the fixed effects $\bmu$, and the variability across individuals, represented by the random effects $\beeta_i$, by combining shared population parameters with subject-specific deviations.

\begin{figure}[H]
\centering

% --- Top-left: PDF plot ---
\begin{subfigure}{0.32\textwidth}
  \centering
  \includegraphics[width=\linewidth]{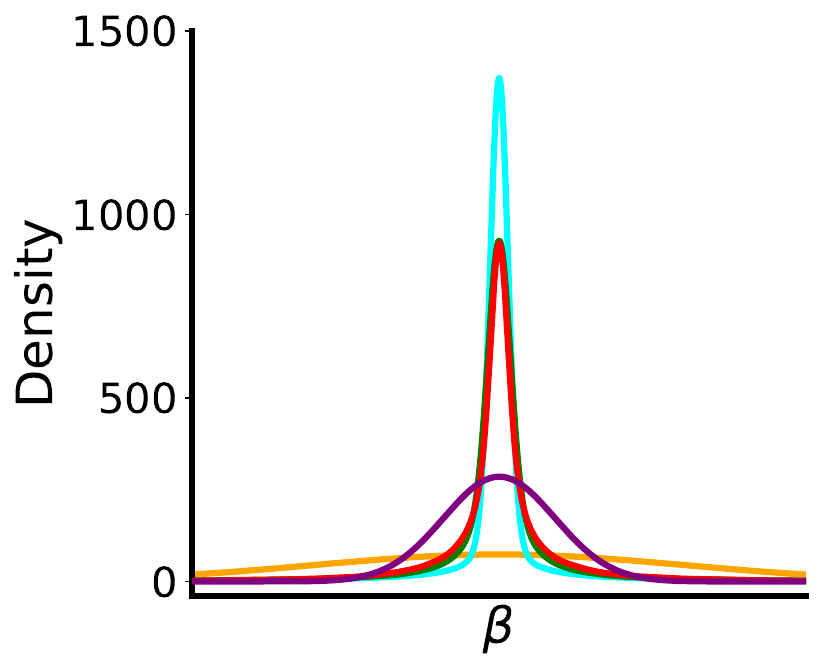} % <- replace filename
  \caption{}
  \label{fig:priors-superposed}
\end{subfigure}
\hfill
% --- Top-right: Table ---
\begin{subfigure}{0.67\textwidth}
  \centering
  \renewcommand{\arraystretch}{1.3}   % vertical padding
  \setlength{\tabcolsep}{2pt}         % horizontal padding
  {\small
\begin{tabular}{lccccc}
\toprule
 & Spike-and-Slab & L1-ball & Hierarchical Lasso & Regularized Horseshoe & R2-D2 \\
\midrule
$\mathbb{P}(|\beta|<10^{-3})$ & 0.972 & 0.823 & 0.751 & 0.798 & 0.853 \\
$\mathbb{P}(|\beta|<10^{-2})$ & 0.975 & 0.841 & 0.993 & 0.986 & 0.986 \\
$\mathbb{P}(10^{-2}<|\beta|<10^{-1})$ & 0.018 & 0.104 & 0.007 & 0.014 & 0.014 \\
$\mathbb{P}(|\beta|>10^{-2})$ & 0.025 & 0.159 & 0.007 & 0.014 & 0.014 \\
$\mathbb{P}(|\beta|>10^{-1})$ & 0.008 & 0.056 & 0.000 & 0.000 & 0.000 \\
$\mathbb{P}(|\beta|>10^{-1}\mid \beta\neq 0)$ & \textcolor{red}{\xmark} & 0.311 & \textcolor{red}{\xmark} & \textcolor{red}{\xmark} & \textcolor{red}{\xmark} \\
\bottomrule
\end{tabular}}
  \caption{}
  \label{fig:prior-table-statistics}
\end{subfigure}

\vspace{0.5em}
\begin{subfigure}{.5\textwidth}
\centering
  \includegraphics[width=1.5\linewidth]{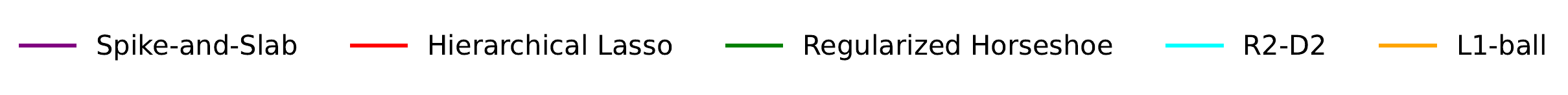}
\end{subfigure}
\vspace{0.75em}
% --- Bottom: single wide PDF plot ---
\begin{subfigure}{\textwidth}
  \centering
  \includegraphics[width=\linewidth]{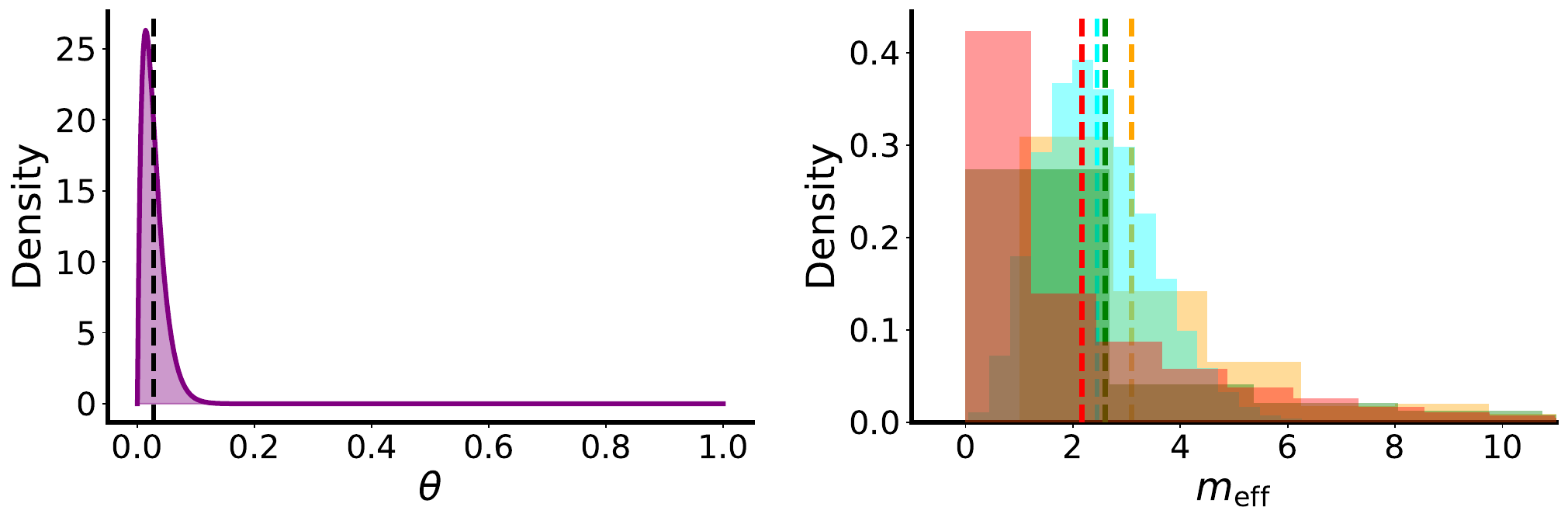}
  \caption{}
  \label{fig:priorchecks}
\end{subfigure}

\caption{(a) Marginal prior distributions for a given SNP effect size $\beta_j$ under the five sparsity-enforcing priors. (b) Prior probabilities of different effect size regimes (with $10^{-3}$ precision), summarizing the mass near zero and in the tails. (c) Left: Beta$(1,30)$ distribution for the global inclusion probability $\theta$ in the spike-and-slab prior. Right: distribution of the effective number of nonzero coefficients $m_{\text{eff}}$ in a SNP effect size vector $\boldsymbol{\beta}$. Kernel density estimates, histograms, and summary statistics are based on 200,000 prior draws.}
\label{fig:fig1}
\end{figure}

Because $K=\sum_{j=1}^p z_j$, from Equation~\ref{eq:sas}, the expected model size is
\begin{equation}
\mathbb{E}[K] = p\mathbb{E}[\gamma] = p\frac{\gamma_a}{\gamma_a+\gamma_b}.
\label{eq:K}
\end{equation}
To target $\tilde m\approx 5$ causal variants out of $p=134$, we set $\mathbb{E}[\gamma]\approx \tilde m/p\approx 0.04$.
For illustration, $\mathrm{Beta}(\gamma_a{=}1,\gamma_b{=}30)$ has mean $1/31\approx 0.032$ (thus $\mathbb{E}[K]\approx 4.3$) and mode at $0$, yielding a stronger bias toward boundary sparsity (Figure~\ref{fig:priorchecks}, left panel).
% If an interior mode near $0.04$ is preferred, choose $\gamma_a,\gamma_b>1$ so that
% %
% \begin{equation}
% \mathrm{mode}(\gamma)=\frac{\gamma_a-1}{\gamma_a+\gamma_b-2}\approx 0.04.
% \end{equation}
% %
For effect-size calibration, we set the slab standard deviation to $\sigma_{\text{slab}}=0.1$.
On the rescaled coefficient scale, this implies
\begin{equation}
p\!\left(|\tilde{\beta}_j|>a \mid z_j=1\right)
=2\left(1-\Phi(a/\sigma_{\text{slab}})\right),
\label{eq:tailss}
\end{equation}
so that, conditional on belonging to the slab component, about $19\%$ of draws exceed $0.13$ in absolute value when $a=0.13$.

% For effect-size calibration, we set the slab standard deviation to $\sigma_{\text{slab}}=0.1$, which assigns non-negligible prior mass to the $H_1$ magnitude $|\beta|=0.13$:
% \begin{equation}
% p\big(|\beta_j|>a \mid z_j{=}1\big)=2(1-\Phi(a/\sigma_{\text{slab}})),
% \label{eq:tailss}
% \end{equation}
% so at $a=0.13$ we get $\approx 0.19$. 

The spike is set to a very small scale, $\sigma_{\text{spike}}=0.0005$ so that
$p(|\beta_j|>0.01\mid z_j{=}0)$ is negligible (derivations in Supplementary Section~\ref{app:ssdetails}).
Summary statistics in Figure~\ref{fig:prior-table-statistics} show that most prior mass lies near zero. Importantly, these table values are computed under the \emph{unconditional} prior of $\beta_j$: they do not condition on $z_j$ but integrate over the spike–slab mixture.

% For the Spike-and-Slab, we choose $(\gamma_a,\gamma_b)$ in Equation~\ref{eq:sas} so that $\mathbb{E}[\gamma]=\gamma_a/(\gamma_a+\gamma_b)\approx \tilde m/p$
% with $\tilde m \approx 5$ and $p=134$, i.e.\ $\mathbb{E}[\gamma]\approx 0.04$.
% Figure~\ref{fig:priorchecks} (left panel) displays a $\mathrm{Beta}(\gamma_a=1,\gamma_b=30)$ distribution,
% whose mean is $1/31\approx 0.032$ (so $\mathbb{E}[m_{\mathrm{eff}}]\approx 4.3$) and whose mode is at $0$, implying a stronger bias toward sparsity at the boundary. 
% If a prior with an interior mode near $0.04$ is desired, one can instead pick $\gamma_a,\gamma_b>1$ with
% $\mathrm{mode}=(\gamma_a-1)/(\gamma_a+\gamma_b-2)$ close to the target.
% For the slab variance we set \(\sigma_{\text{slab}}=0.1\), which allocates non–negligible mass to the effect size observed under $H_1: p(|\beta_j|>0.13\mid z_j{=}1) \approx 0.19$.
% The spike variance is \(\sigma_{\text{spike}}^2=(0.0005)^2\) so that
% $p(|\beta_j|>0.01\mid z_j{=}0)$ is negligible (analytical derivations in Supplementary Section~\ref{app:ssdetails}).
% Summary statistics in Figure~\ref{fig:prior-table-statistics} show that most prior mass lies near zero. Importantly, these table values are computed under the \emph{unconditional} prior of $\beta_j$: they do not condition on $z_j$ but integrate over the spike–slab mixture.

\underline{$\ell_1$--ball prior calibration.}

We proceed in two stages, calibrating first the expected magnitude of causal SNPs and subsequently the expected number of causal SNPs.
With the projection involving a common soft-threshold, the memoryless property implies that, conditional on being active,
$|\tilde\beta_j| \mid (\tilde\beta_j \neq 0)\sim \mathrm{Exp}(\text{rate}=1/b_\xi),
$
and therefore
$\mathbb{E}\!\left[|\tilde\beta_j| \mid \tilde\beta_j \neq 0\right]=b_\xi$.
With the projection involving a common soft--threshold, the memoryless property implies that, conditional on being active,
\begin{equation}
|\tilde\beta_j|
\ \stackrel{d}{=}\
|\tilde\xi_j|-\tilde\delta\ \big|\ |\tilde\xi_j|>\tilde\delta
\ \sim\ \mathrm{Exp}(\text{rate}=1/b_\xi).
\label{eq:memoryless}
\end{equation}
Equivalently, on the original scale, $|\beta_j|\mid(\beta_j\neq 0)$ has mean $\omega_h b_\xi$ and tail
\[
p\big(|\beta_j|>a\ \big|\ \beta_j\neq 0\big)=\exp\!\left(-\frac{a}{\omega_h b_\xi}\right).
\]
% We set $b_\xi$ to match the desired effect scale, e.g.\ by targeting
% $p(|\beta_j|>0.13\mid \beta_j\neq 0)\approx 0.27$.
We set $b_\xi$ to match the desired effect scale, for example by targeting
a conditional tail probability $p(|\beta_j|>0.13 \mid \beta_j\neq 0)$ in a
comparable range (roughly $0.2$--$0.3$ in our calibration).

Next, sparsity is governed by the $\ell_1$ budget $\tilde r\sim \mathrm{Exp}(\lambda_r)$.
Since the projection enforces
$\sum_{j:\tilde\beta_j\neq 0} |\tilde\beta_j|=\tilde r$, and each active $|\tilde\beta_j|$ has mean $b_\xi$
(Equation~\ref{eq:memoryless}), the expected number of active SNPs is approximated by
\begin{equation}
\mathbb{E}[K] \approx \frac{\mathbb{E}[\tilde r]}{b_\xi} = \frac{1}{\lambda_r b_\xi}.
\end{equation}
Given $b_\xi$, we then choose $\lambda_r$ so that $\mathbb{E}[K]\approx \tilde m \approx 5$,
i.e.\ the prior expects only a handful of nonzero SNP effects among the $p=134$ candidates.

\underline{Global–local shrinkage priors.}
The hierarchical lasso, regularized horseshoe and R2D2 baselines all employ a global–local Gaussian scale–mixture prior (Equations~\ref{eq:betahl},~\ref{eq:betahs} and~\ref{eq:betar2}), which do not yield exact zeros. 
To formalize the notion of model size, we follow~\cite{Piironen_2017} and define it through the expected shrinkage applied to ordinary least–squares. A full derivation can be found in Supplementary Section~\ref{app:shrinkagefactor}. 
%We adopt standard notations and refer to the response modeled by genetic variations, the log-clearance, as $\y$.

Under the orthogonal-design Gaussian proxy of~\cite{Piironen_2017}, used here \emph{only} for prior calibration,
we center and scale the SNP columns so that each has unit variance and $\|\mathbf{x}_j\|_2^2=N$.
Genetic effects enter the PK model through the latent log-clearance.
The residual interindividual variability on log-clearance, $\omega_{h}^2$ (Eq.~\ref{eq:randomeffects}),
therefore sets the natural scale of the regression noise in this proxy.
In this setting, the ordinary least-squares signal for coordinate $j$ is
$\hat\beta_j=(\mathbf{x}_j^\top \mathbf{y})/N$ and the coordinate-wise likelihood decouples as
\begin{equation}
\mathcal{L}_j(\beta_j) \propto \exp\!\left(-\frac{N}{2\omega_h^2}(\beta_j-\hat\beta_j)^2\right).
\label{eq:coordlike}
\end{equation}
which is Gaussian with mean $\hat\beta_j$ and precision $N/\omega_{h}^2$.

For global–local shrinkage priors, the regression coefficients admit the generic representation
\begin{equation}
\beta_j \mid \tau, w_j
\sim
\mathcal N\!\big(0,\omega_h^2 \tau^2 w_j\big),
\label{eq:generic_gl_prior}
\end{equation}
where $\tau$ is a global scale parameter and $w_j$ are prior-specific local weights.
The definition of $w_j$ depends on the chosen prior:
\begin{equation}
w_j =
\begin{cases}
\lambda_j^{2}
& \text{Hierarchical Lasso (Eq.~\ref{eq:betahl})},\\[4pt]
\tilde{\lambda}_j^{2}
& \text{Regularized Horseshoe (Eq.~\ref{eq:betahs})},\\[4pt]
\lambda_j
& \text{R2--D2 (Eq.~\ref{eq:betar2})}.
\end{cases}
\label{eq:w_def}
\end{equation}

Combining~\eqref{eq:coordlike} and~\eqref{eq:generic_gl_prior} yields a conjugate Gaussian update.
The posterior mean takes a ridge-type form:
\begin{equation}
\mathbb{E}[\beta_j \mid \mathbf{y},\tau,w_j]
=
(1-\kappa_j)\hat\beta_j,
\qquad
\kappa_j
=
\frac{1}{1 + N\tau^2 w_j}.
\label{eq:kappa}
\end{equation}
Importantly, because both the likelihood~\eqref{eq:coordlike} and the prior~\eqref{eq:generic_gl_prior}
are expressed in units of $\omega_h^2$, the shrinkage factor $\kappa_j$ in~\eqref{eq:kappa}
does not depend on the overall scale of the phenotype.
Calibration therefore depends only on $(N,p)$ and the desired sparsity level,
but is invariant to the magnitude of the pharmacokinetic parameter itself.
Here $\kappa_j$ is the posterior shrinkage factor: $\kappa_j\approx 1$ means $\hat\beta_j$ is fully suppressed, $\kappa_j\approx 0$ means it is left essentially unshrunk.

This univariate analysis is justified by the orthogonal design approximation, which decouples coordinates so that the posterior for each $\beta_j$ can be studied in isolation.
We define the \emph{effective prior model size} as
\begin{equation}
m_{\text{eff}}=\sum_{j=1}^p (1-\kappa_j),
\end{equation}
which is the summary used by \cite{Piironen_2017} to calibrate the global scale via a target sparsity.
When $\kappa_j$’s concentrate near 0 or 1, as they typically do under heavy–tailed global–local priors, $m_{\mathrm{eff}}$ counts how many coefficients are effectively unshrunk, i.e.\ how many predictors the prior expects to be \emph{active} or how many variants the prior expects to be causal in our case.
This provides the calibration target: by choosing the global scale so that $\mathbb{E}[m_{\mathrm{eff}}]\approx \tilde m$, we translate sparsity assumptions into hyperparameter values. These are presented in Equation~\ref{eq:hpval}, with the complete derivation provided in Supplementary Section~\ref{app:hpval}:
\begin{equation}
\begin{cases}
\tau_{0} \approx \dfrac{\rho}{\sqrt{2N}}\sqrt{\dfrac{\tilde m}{p-\tilde m}}
& \text{Hierarchical Lasso,}\\[12pt]
\tau_{0} \approx \dfrac{1}{\sqrt{N}}\dfrac{\tilde m}{p-\tilde m} 
& \text{Regularized Horseshoe,}\\[10pt]
\mathbb{E}[R^{2}] \approx 
\dfrac{\displaystyle \tfrac{p}{N}\tfrac{\tilde m}{p-\tilde m}}
{1+\displaystyle \tfrac{p}{N}\tfrac{\tilde m}{p-\tilde m}}
& \text{R2--D2}.
\end{cases}
\label{eq:hpval}
\end{equation}
Lastly, magnitude calibration proceeds by tuning the remaining hyperparameters (slab width, $\rho$, $R^2$ target)
to place a desired prior mass on clinically meaningful effect sizes under $H_1$
(e.g.\ $p(|\beta|>0.13)\in[0.15,0.30]$),
which under our scale-invariant parameterization corresponds to
$p\!\left(|\tilde\beta|>\frac{0.13}{\omega_h}\right)\in[0.15,0.30]$.
We finally validate both $\mathbb{E}[m_{\mathrm{eff}}]$ and the tail probabilities by prior Monte Carlo.

Except for the Spike-and-Slab, we can get Monte Carlo draws of the effective prior number of active SNPs ($K$ for the $\ell_1$-ball, $m_{\text{eff}}$ for the global-local shrinkage priors). These are displayed in Figure~\ref{fig:priorchecks} (right panel). Finally, Figure~\ref{fig:priors-superposed} shows all the priors overlaid.

For all approaches, a practical implementation is provided using the probabilistic programming language NumPyro~\citep{phan2019composable, bingham2019pyro}, see Supplementary Section~\ref{sec:pseudocodes}.
Table~\ref{tab:prior_hyperparameters} provides the hyperparameters of each prior.

% Additional details can be found in Supplementary Section~\ref{app:.

\subsection{Metrics}\label{sec:metrics}

Throughout the paper, we assess variable-selection performance using both Bayesian posterior summaries and classical error-control metrics. First, we compute \emph{posterior inclusion probabilities} for each SNP and obtain selections by applying a decision cutoff to the PIPs. Second, to situate results within a multiple-testing framework, we report the \emph{family-wise error rate} under $H_0$ (no true signals) and the $F_1$-score under $H_1$ (averaged over datasets), which balances precision and recall. These latter metrics are induced by PIPs, which are standard in Bayesian variable selection and model averaging~\citep{barbieripip,stephenspip,finemappip}.

\paragraph{Posterior inclusion probabilities (PIPs).}
For SNP $j\in\{1,\dots,p\}$ and the $d$th dataset $\mathcal D^{(d)}$, we define:
\begin{align}
\textbf{Analytical PIP (spike--slab):}\quad
\pi_{j}^{(d)}
&:= p\!\big(z_j=1 \mid \mathcal D^{(d)}\big)
= \mathbb{E}\!\left[p\!\big(z_j=1\mid \beta_j,\gamma\big)\middle|\mathcal D^{(d)}\right]
\approx \frac{1}{S}\sum_{s=1}^{S} p\!\big(z_j=1\mid \beta_j^{(s)},\gamma^{(s)}\big),
\label{eq:analytical-pip}
\\
\textbf{Proxy PIP (generic):}\quad
\widehat{\pi}_{j}^{(d)}(\varepsilon)
&:= p\!\big(|\beta_j|>\varepsilon \mid \mathcal D^{(d)}\big)
\approx
\frac{1}{S}\sum_{s=1}^{S}\mathds{1}\!\left\{ \left|\beta_{j}^{(s)}\right|>\varepsilon \right\},
\label{eq:proxy-pip}
\end{align}
where $z_j\in\{0,1\}$ is the spike--slab inclusion indicator, 
$\{\bbeta^{(s)}\}_{s=1}^S$ are posterior draws, and $\varepsilon>0$ is a small effect-size tolerance.
For the spike-and-slab prior, we use Equation~\ref{eq:analytical-pip}, obtained by evaluating the closed-form conditional inclusion probability \(p(z_j=1\mid \beta_j^{(s)},\gamma^{(s)})\) at each posterior draw and averaging over draws. This process is known as Rao--Blackwellization and replaces the latent Bernoulli indicator by its conditional expectation, thereby reducing Monte Carlo variance.

\paragraph{Family-Wise Error Rate (FWER).}
Let $\mathbb{D}_{0}$ be the set of datasets generated under $H_0$ (no truly associated SNPs).
Given a PIP inclusion cutoff $\tau\in(0,1)$, we declare SNP $j$ \emph{selected} in dataset $\D^{(d)}$ if 
$\text{PIP}_j^{(d)}>\tau$.
The family-wise error rate is estimated as the fraction of $H_0$ datasets with at least one selection:
\begin{equation}
\widehat{\text{FWER}}(\tau,\varepsilon)
=
\frac{1}{|\mathbb{D}_{0}|}
\sum_{d\in\mathbb{D}_{0}}
\mathds{1}\!\left\{\max_{1\le j\le p}\text{PIP}_j^{(d)}(\varepsilon)>\tau\right\}.
\label{eq:fwer}
\end{equation}
We report a binomial Wilson $95\%$ confidence interval for Equation~\ref{eq:fwer} across $\mathbb{D}_0$.

\paragraph{$F_1$-score.}
Let $\mathbb{D}_{1}$ be the set of $H_1$ datasets and $\mathcal{P}^\star\subseteq\{1,\dots,p\}$ the set of truly associated SNPs.
Given $\tau$ and $\varepsilon$, define the selected set in dataset $d$ as
\[
\widehat{\mathcal{P}}^{(d)}(\tau,\varepsilon)
=
\big\{ j:\ \text{PIP}_j^{(d)}(\varepsilon)>\tau \big\}.
\]
Per-dataset precision and recall are
\[
\text{Prec}^{(d)}=\frac{|\widehat{\mathcal{P}}^{(d)}\cap \mathcal{P}^\star|}{|\widehat{\mathcal{P}}^{(d)}|\vee 1},
\qquad
\text{Rec}^{(d)}=\frac{|\widehat{\mathcal{P}}^{(d)}\cap \mathcal{P}^\star|}{|\mathcal{P}^\star|},
\]
and the per-dataset F$_1$-score is
\[
F_1^{(d)}=\frac{2\text{Prec}^{(d)}\text{Rec}^{(d)}}{\text{Prec}^{(d)}+\text{Rec}^{(d)}}\ \ \in[0,1].
\]
We report the average score over $H_1$ datasets,
\begin{equation}
\overline{F_1}(\tau,\varepsilon)
=
\frac{1}{|\mathbb{D}_{1}|}\sum_{d\in\mathbb{D}_{1}}F_1^{(d)},
\label{eq:f1}
\end{equation}
with a standard error across $\mathbb{D}_{1}$.

Unless stated otherwise, we run 5 independent NUTS chains and retain 2000 post-warm-up draws per chain. Convergence and mixing are assessed with standard diagnostics: split $\widehat{R}\!<\!1.01$, bulk and tail effective sample sizes per parameter. For each of the two hypotheses (\(H_0\) and \(H_1\)), we generated 100 independent simulated datasets and ran all five priors on each replicate.

\subsection{Results}

\paragraph{Posterior inference for PK parameters and convergence.}
Across both hypotheses, posterior distributions for the core PK parameters and variance components are largely consistent across priors and concentrate around the ground-truth values used to generate the data (Figure~\ref{fig:kde}).
This indicates that introducing sparsity on the genetic effects does not materially distort inference on the pharmacokinetic component of the model.
As expected, the posterior for the causal SNP effect (denoted $\beta_{\text{rs37}}$ in the figures) concentrates near $0$ under $H_0$, while under $H_1$ it displays a mixture of mass near the true effect and near $0$ when aggregated over datasets (Figure~\ref{fig:kde}), reflecting the fact that some simulated datasets provide stronger evidence for the signal than others.
MCMC diagnostics are overall satisfactory across the 100 replicated datasets, with most split-$\widehat{R}$ values remaining below the common $1.1$ threshold for the main PK parameters and variance components (Figure~\ref{fig:rhat}).

\paragraph{Type-I error under $H_0$.}
We next evaluate false-discovery control using the family-wise error rate (FWER) induced by the posterior inclusion probabilities defined in Section~\ref{sec:metrics} (with proxy PIPs evaluated at the default tolerance $\varepsilon=0.01$ when needed).
Using a PIP cutoff $\tau=0.4$ to declare discoveries, Table~\ref{tab:ss-top5-analytical-H0} shows that all methods are conservative or close to nominal in this setting: the spike-and-slab achieves $\mathrm{FWER}=0.040$ (Wilson 95\% CI $[0.016,0.098]$), the regularized horseshoe yields $0.020$ ($[0.006,0.070]$), and the hierarchical lasso and R2--D2 exhibit no false rejections in the 100 replicates (CIs up to $0.037$).
The $\ell_1$-ball prior is comparatively more liberal at this cutoff
($\mathrm{FWER}=0.080$, CI $[0.041,0.150]$), indicating that under $H_0$
it more often yields at least one null SNP above the decision threshold.
% The $\ell_1$-ball prior is comparatively more liberal at this cutoff ($\mathrm{FWER}=0.080$, CI $[0.041,0.150]$), consistent with its comparatively larger proxy inclusion probabilities under $H_0$ (Fig.~\ref{fig:inclprob}, top row).

\begin{figure}[H]
    \centering
    \includegraphics[width=1\linewidth]{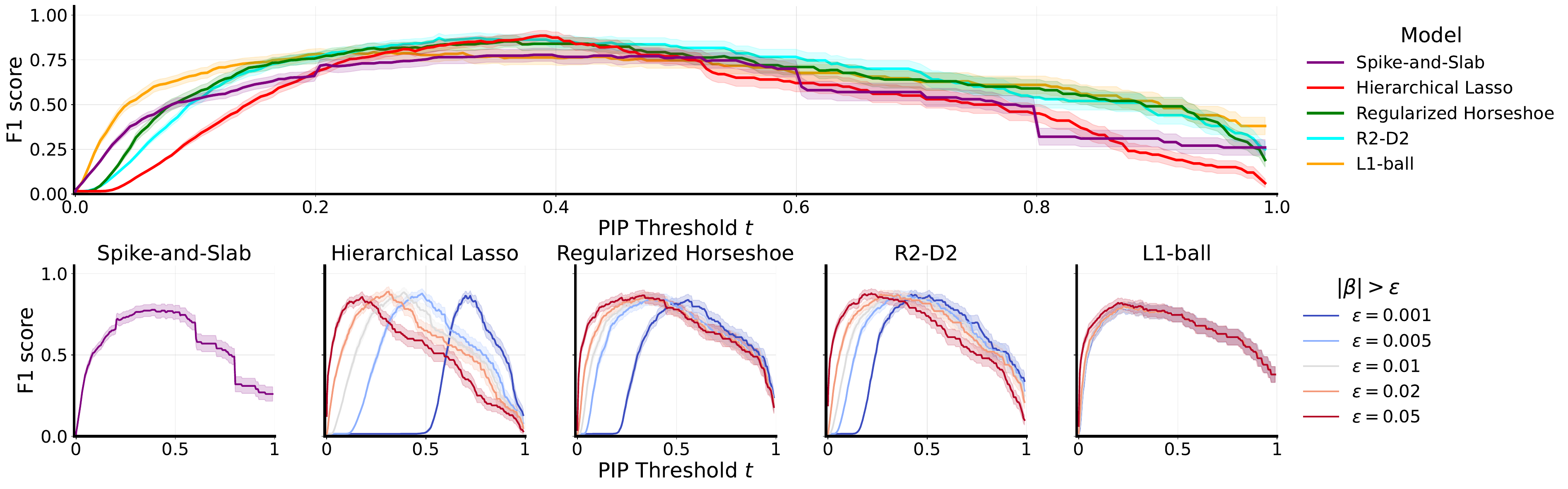}
    \caption{$F_1$-score as a function of the posterior inclusion probability (PIP) threshold. Top panel: comparison of all five sparsity-inducing priors at fixed $\varepsilon=0.01$. Bottom row: one panel per prior, showing $F_1$ curves across multiple values of $\varepsilon$, except for the spike-and-slab, where analytical PIPs were used. Shaded regions denote standard errors computed over 100 datasets.}
    \label{fig:fig3}
\end{figure}

\paragraph{Detection performance under $H_1$ and choice of the PIP cutoff.}
Under $H_1$, we summarize variable-selection accuracy using the $F_1$-score induced by PIP thresholding.
Figure~\ref{fig:fig3} (top panel, $\varepsilon=0.01$) shows that all priors reach a broad high-performance plateau, with $F_1$ peaking for intermediate PIP cutoffs (roughly $\tau\in[0.2,0.6]$).
This robustness motivates using a single default cutoff (here $\tau=0.4$) for reporting error-control metrics while remaining close to the empirical optimum across priors.
Differences between priors are modest near the peak, but appear more pronounced in the low-threshold regime: the $\ell_1$-ball and global--local shrinkage priors tend to achieve higher $F_1$ at small $\tau$, while the hierarchical lasso requires a more stringent cutoff to avoid degrading precision.

\paragraph{Sensitivity to the tolerance parameter $\varepsilon$ for proxy PIPs.}
For priors that do not provide inclusion probabilities directly, the tolerance $\varepsilon$ controls what constitutes a practically non-negligible effect in the proxy PIP.
Figure~\ref{fig:inclprob} supports the default choice $\varepsilon=0.01$: at this value, the causal SNP remains clearly separated from non-causal SNPs under $H_1$, while null proxy-inclusion probabilities under $H_0$ remain close to zero.
For smaller tolerances, null proxy inclusion increases under $H_0$, which can reduce precision, whereas larger tolerances can down-weight moderate effects and reduce power.
Figure~\ref{fig:fig3} (bottom row) shows that the induced $F_1$ curves are generally stable around $\varepsilon=0.01$, although the Hierarchical Lasso appears more sensitive to $\varepsilon$ than the other proxy-PIP baselines.
By contrast, the $\ell_1$-ball appears comparatively robust across the range of $\varepsilon$ values considered.
This is consistent with its projection-based soft-thresholding mechanism, which sets many coefficients exactly to zero, so moderate changes in $\varepsilon$ near zero have limited effect on the resulting proxy-PIP summaries.
Overall, these results support using $\varepsilon=0.01$ as a simple, stable default for proxy-PIP comparisons in the simulation study, while retaining analytical PIPs for the spike-and-slab prior.

\section{Real case study}\label{sec:realstudy}

\subsection{Study design.}

The aims of the ANRS12154 open-label, single-center, multiple-dose PK study were to characterize nevirapine PK in a Cambodian population living with HIV and to identify
environmental and genetic factors of variability.
A total of 170 Cambodians living with HIV were included.
Nevirapine trough concentrations were measured
after 18 and 36 months of starting antiretroviral treatment and in samples drawn during a dosing
interval in a subset of 10 patients (predose
and 1 h, 2 h, 4 h, and 8 h after the nevirapine morning intake).
These concentrations were analyzed using a nonlinear
mixed-effect model approach with MONOLIX software version 2.4 (http://
software.monolix.org/). 
The model-building process is detailed elsewhere~\cite{Chou2010NevirapinePK}. 
A one compartment model with first-order absorption and elimination adequately described nevirapine concentrations. 
Using this population model, empirical Bayes estimates of individual nevirapine clearances at months 18 and 36 were derived for the 129 patients with available genetic data.
We calculated the average over the two (three for the 10 patients also with a complete pharmacokinetic profile) estimates to be used as the phenotype
for genetic association analyses. One individual outlier (extremely high) nevirapine clearance value was censored from the analyses. From the ANRS12154 cohort, targeted genotype data were available for 198 SNPs in 7 genes (47 in CYP2B6, 1 in CYP2A6, 1 in CYP2C19, 63 in ABCB1, 36 in CYP3A4, 1 in CYP3A5, and 49 in NR1I2)~\cite{Bertrand2012}. 

Here, the clearance model described in Equation~\ref{eq:lrcl} is extended to include between-occasion random effects.
Additionally, the well-established pharmacogenetic variant \texttt{rs3745274} (CYP2B6 516G→T), known to reduce nevirapine clearance, is not
modeled as a separate covariate: it is included in the SNP design matrix and is therefore
subject to the same sparsity-inducing prior as the other genetic effects~\cite{Bertrand2012}. This provides an
internal validity check: a sensible sparsity prior should recover \texttt{rs3745274} while still
allowing for additional (weaker) polygenic signals.

Let $o$ index occasions for subject $i$. With standardized genotypes $X_{ij}$ for $p$ SNPs,  we use
\begin{equation}
\label{eq:lrcl_bov}
\log CL_{i,o}
= \log \hat{CL}
+ \sum_{j=1}^{p} X_{ij}\beta_j
+ \eta_{i,CL}
+ \kappa_{i,o,CL}.
\end{equation}
Compared to Equation~\ref{eq:lrcl}, we add a between-occasion (within-subject) random effect
$\kappa_{i,o, CL}$ shared by all observations in occasion $o$, with
\begin{equation}
\kappa_{i,o} \sim \mathcal{N}(0,\psi_{CL}^2), \qquad \psi_{CL} \sim \mathcal{H}\mathcal{N}(0.5),
\end{equation}
thereby placing a tighter prior on the intra-occasion scale to encode that within-subject
variability is smaller than between-subject variability (Equation~\ref{eq:interindivvar}).

\paragraph{Normalized Prediction Discrepancy (NPD)~\citep{vpc}.}
NPDs are useful visual predictive checks that map each observation to its posterior–predictive CDF and then to a standard–normal score via $\Phi^{-1}$. Under a well–calibrated model these scores are approximately $\mathcal{N}(0,1)$. Systematic deviations (shifted median, overly wide/narrow spread, or time trends) reveal bias, mis–specified variability, or time–dependent lack of fit.

To compute NPDs, we draw $S$ posterior-predictive replicates for each observed point:
\begin{equation}
\tilde y_{i}^{(l,s)} = f\!\big(\boldsymbol{\phi}_i^{(s)}; t_i^{(l)}, D\big) + v_i^{(l,s)},
\qquad
\boldsymbol{\phi}_i^{(s)} = \exp\!\big(\boldsymbol{\mu}^{(s)} + \boldsymbol{\eta}_i^{(s)}\big),
\qquad s=1,\dots,S,
\end{equation}
with $(\boldsymbol{\mu}^{(s)},\boldsymbol{\eta}_i^{(s)})$ and $v_i^{(l,s)}$ sampled from their posterior and observation-noise distributions, respectively.
Let $R_i^{(l)}$ be the rank of the observed $y_i^{(l)}$ among the $S$ replicates:
\begin{equation}
R_i^{(l)} = \sum_{s=1}^{S} \mathds{1}\!\left\{\tilde y_i^{(l,s)} < y_i^{(l)}\right\},
\qquad
\hat p_i^{(l)} = \frac{R_i^{(l)} + 0.5}{S + 1}.
\end{equation}
The normalized prediction discrepancy is
\begin{equation}
\text{NPD}_i^{(l)} = \Phi^{-1}\!\big(\hat p_i^{(l)}\big),
\end{equation}
which should be approximately standard normal under a well-calibrated model. In diagnostics, we compare observed time-binned percentiles of $\{\text{NPD}_i^{(l)}\}$ (e.g., 10th/50th/90th) against the corresponding theoretical quantiles and their sampling-uncertainty bands.

\paragraph{Prior calibration on real data.}
In the 12154 ANRS analysis, we expect the dominant genetic contribution to clearance to arise from \texttt{rs3745274}, whose effect on the log-clearance scale is of order $10^{-1}$, together with at most a handful of additional SNPs with smaller effects (typically $10^{-2}$ to $5\times 10^{-2}$)~\cite{Bertrand2012}.
We therefore calibrate the sparsity priors to concentrate most mass near zero while retaining sufficient tail mass to accommodate one comparatively large coefficient and a few moderate signals. The obtained prior statistics for this setting are display in Figure~\ref{fig:fig1real}. \\

We run 5 independent NUTS chains with 3000 warm-up iterations and 2000 post-warm-up draws per chain.
Unless stated otherwise, posterior inclusion probabilities for continuous shrinkage priors are reported at tolerance $\varepsilon=0.01$ using the proxy $\mathbb{P}(|\beta_j|>\varepsilon\mid \mathcal{D})$, while for Spike-and-Slab we use analytical inclusion probabilities $p(z_j=1\mid\mathcal{D})$.
\subsection{Results}

\subsubsection{Pharmacokinetic parameter inference and model calibration}

\begin{figure}[H]
    \centering
    \includegraphics[width=1\linewidth]{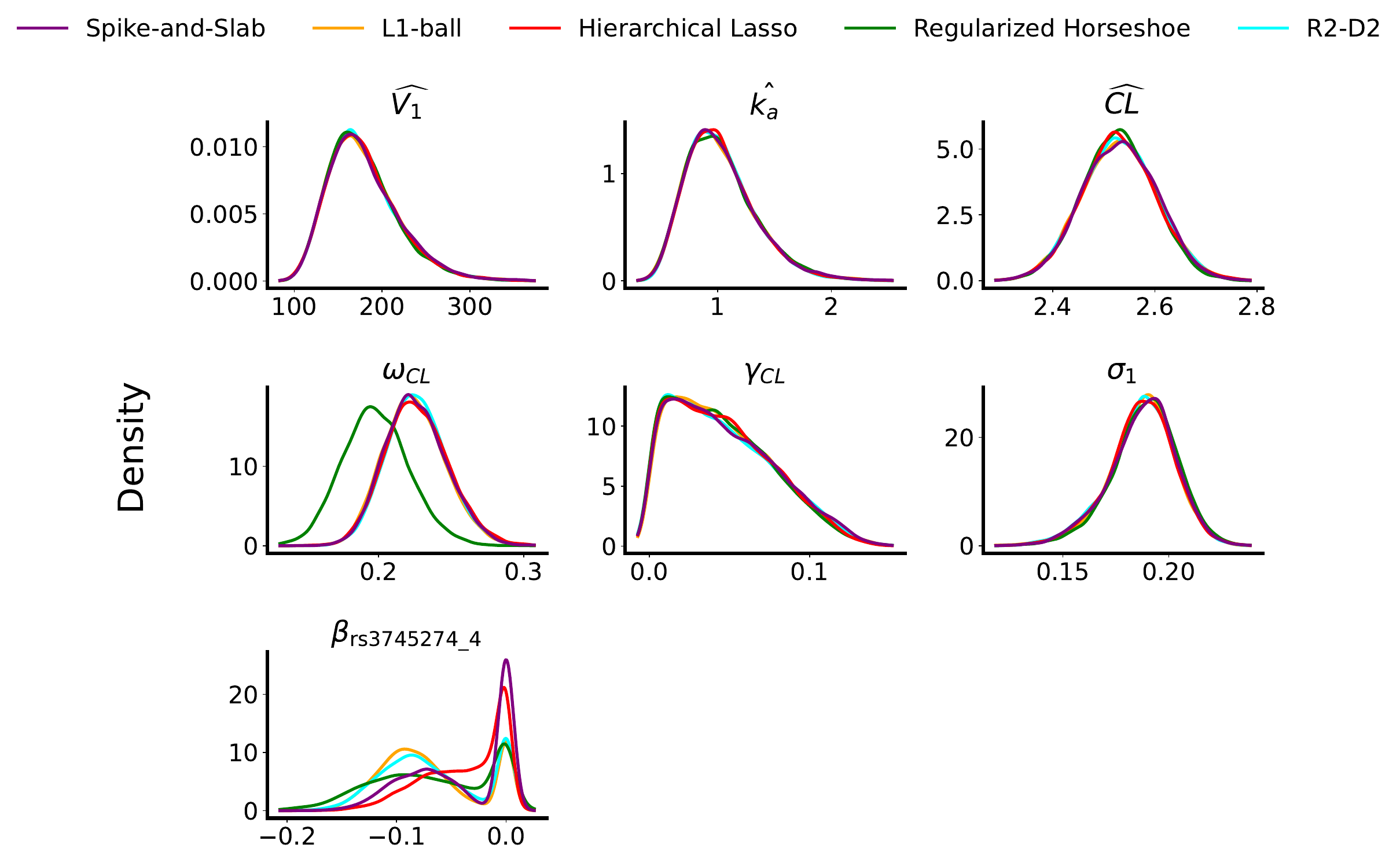}
    \caption{Posterior distributions of pharmacokinetic parameters across the five sparsity-inducing priors. Shown are the population parameters, inter- and intra-individual variability terms, the residual error, and the SNP effect $\beta_{\mathrm{rs3745274}}$. Kernel density estimates are based on draws from 5 NUTS chains with 2000 samples each.}
    \label{fig:kde_real}
\end{figure}

\paragraph{PK parameters and between-/within-subject variability.}
Across priors, posterior distributions of the population parameters $(\widehat{\mu_{V_1}},\widehat{\mu_{k_a}},\widehat{\mu_{CL}})$
and the residual error $\widehat{\sigma_1}$ are highly consistent (Figure~\ref{fig:kde_real}), indicating that the
choice of sparsity prior on genetic effects has limited impact on core PK inference in this dataset.
Posterior uncertainty for the within-subject (between-occasion) term $\gamma_{CL}$ remains similarly stable.
However, the Regularized Horseshoe yields a notably shifted posterior for the inter-individual clearance scale $\omega_{CL}$ compared to the other priors (Figure~\ref{fig:kde_real}). One possible explanation is a variance-allocation trade-off: because heavier-tailed shrinkage allows
a small number of genetic coefficients to remain less shrunk, part of the between-subject variation in clearance may be absorbed by these effects rather than by the residual variability term $\omega_{CL}$.

\begin{figure}[H]
    \centering
    \includegraphics[width=1\linewidth]{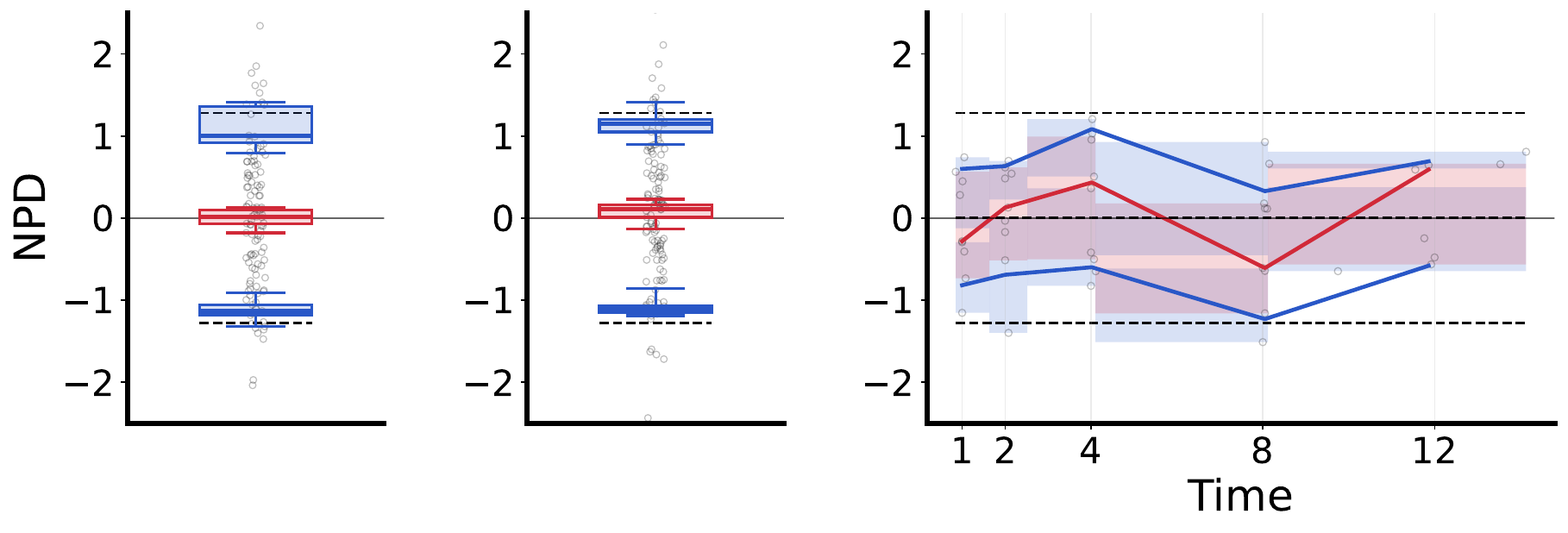}
    \caption{Normalized prediction distribution (NPD) visual predictive checks by sampling scheme. The sparsity-enforcing prior used here is the $\ell_1$-ball. Left and middle panels: subjects with a single sampling occasion (occ = 3 and occ =4). Each panel shows three boxplots—blue for the 10th and 90th percentiles, red for the median—summarizing the bootstrap sampling distribution of the corresponding observed NPD percentiles; solid colored bars mark the observed percentiles themselves, dashed black lines mark the standard normal reference percentiles. Open circles are individual NPD values for each observation, horizontally jittered to avoid clutter. Right panel: subjects with multiple samples (occ = 1). Points are NPDs versus time; colored lines are the observed 10th, 50th, and 90th percentiles across quantile-based time bins, with matching shaded 95\% bootstrap confidence bands; dashed black lines are the standard normal reference percentiles. NPDs are computed from the predictive distribution using mid-rank probabilities from $K$ simulated replicates per panel. Consistency of colored summaries with the dashed references indicates adequate calibration; systematic deviations suggest model misfit (e.g., bias over time).}
    \label{fig:npd}
\end{figure}

\paragraph{Predictive calibration.}
We assessed predictive adequacy using NPD checks (Figure~\ref{fig:npd}), shown here for the L1-Ball prior.
Across sampling schemes (single-occasion and multiple-occasion individuals), observed NPD summaries (10th/50th/90th percentiles) broadly align with the $\mathcal{N}(0,1)$ reference values, with no pronounced systematic drift over time in the multi-sample panel.
Overall, these diagnostics suggest that the model (including between-occasion effects) provides a reasonable description of the concentration--time data.

\subsubsection{Genetic signal recovery}

\begin{figure}[H]
    \centering
    \includegraphics[width=1\linewidth]{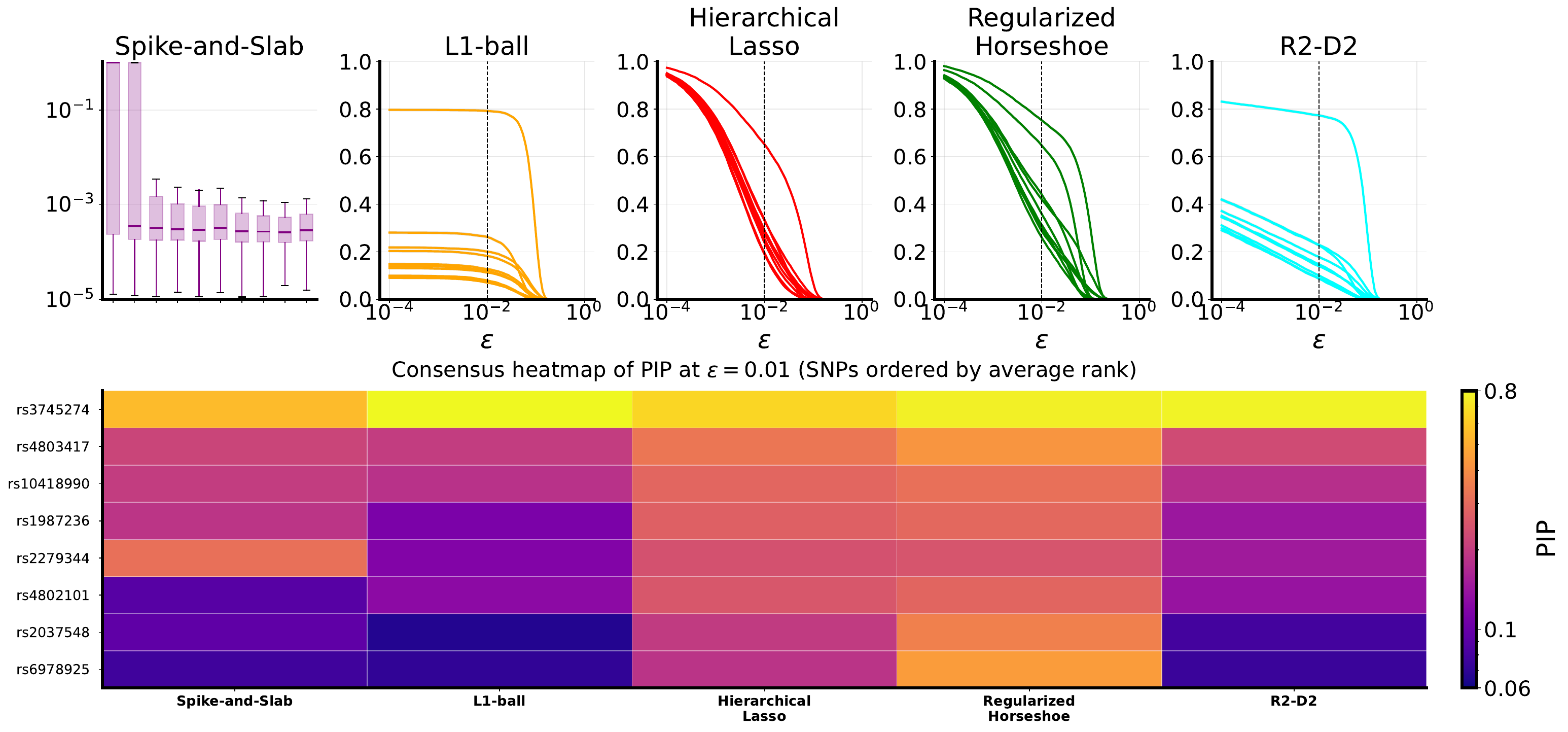}
\caption{
Posterior inclusion probabilities across sparsity-inducing priors for the real data case study.
\textbf{Top row:} For the continuous shrinkage priors (Hierarchical Lasso, Regularized Horseshoe, R2-D2, L1-ball), curves show the inclusion probability $\mathbb{P}(|\beta_j| > \varepsilon)$ for the top-ranked SNPs as a function of the effect-size tolerance $\varepsilon$ (log scale). 
For the Spike-and-Slab prior, boxplots display the posterior distribution of analytical inclusion probabilities $p(z_j = 1 \mid \mathcal{D})$ across MCMC draws.
Vertical dashed lines mark $\varepsilon = 0.01$.
\textbf{Bottom row:} Consensus heatmap of posterior inclusion probabilities at $\varepsilon = 0.01$. 
Rows correspond to SNPs ordered by their average rank across priors, and columns correspond to priors. 
Color intensity (log scale) represents the magnitude of the posterior inclusion probability. 
All posterior summaries are computed from 5 independent NUTS chains with 2000 post-warm-up samples per chain.
}
    \label{fig:summary}
\end{figure}

\begin{figure}[H]
    \centering
    \includegraphics[width=1\linewidth]{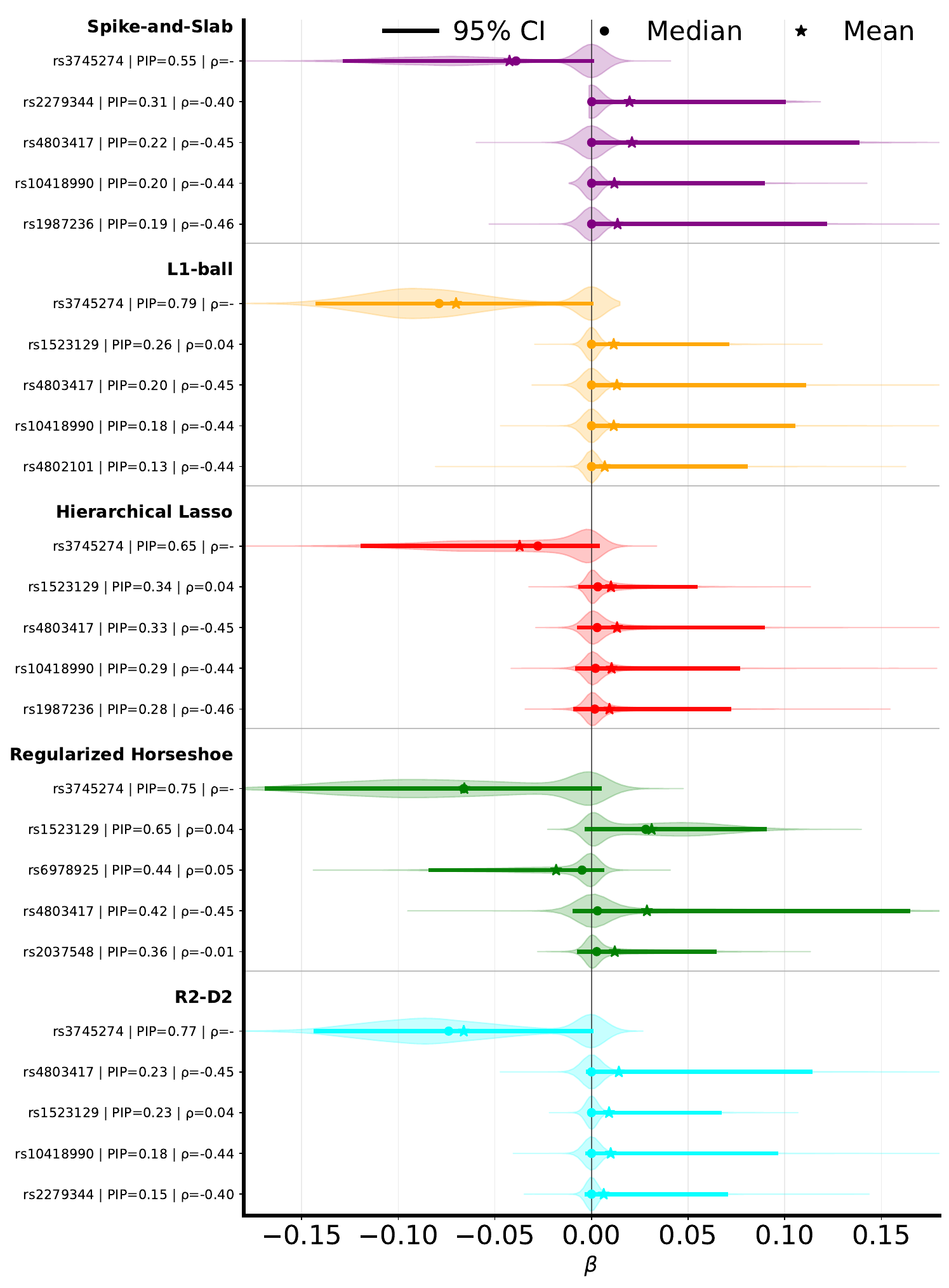}
\caption{
Posterior distributions of the top five SNP effects per sparsity-inducing prior in the real data case study.
For each prior, the five horizontal violins correspond to the SNPs with the highest posterior inclusion probability (PIP) under that prior.
Row labels report the SNP identifier, its PIP at $\varepsilon = 0.01$, and its Pearson correlation $\rho$ with the known variant \texttt{rs3745274}.
Violin shapes depict the full posterior distribution of $\beta$; thick horizontal segments indicate 95\% posterior credible intervals, circles mark posterior medians, and stars denote posterior means.
}
    \label{fig:violin}
\end{figure}

\paragraph{Recovery of \texttt{rs3745274}.}
Figure~\ref{fig:summary} summarizes inclusion behavior across priors. For continuous shrinkage priors, the top-row curves report the proxy
$\mathbb{P}(|\beta_j|>\varepsilon \mid \mathcal{D})$ as a function of the tolerance $\varepsilon$, whereas for
Spike-and-Slab we report analytical inclusion probabilities $p(z_j=1\mid\mathcal{D})$ obtained by
Rao--Blackwellization.
Across all five priors, \texttt{rs3745274} is clearly separated from the remaining SNPs: its inclusion
support remains substantially higher over a wide range of $\varepsilon$, with the strongest persistence under the
$\ell_1$-ball and a similar, though more strongly shrunk, pattern under the Regularized Horseshoe.
At $\varepsilon=0.01$, its inclusion probability ranges from $0.55$ (Spike-and-Slab) to $0.79$ ($\ell_1$-ball),
with intermediate values for the Hierarchical Lasso ($0.65$), Regularized Horseshoe ($0.75$), and R2--D2 ($0.77$)
(Figure~\ref{fig:violin}).
Consistently, the posterior for $\beta_{\mathrm{rs3745274}}$ is shifted away from zero, with effect size of order
$10^{-1}$ on the log-clearance scale (Figure~\ref{fig:kde_real}), in line with \texttt{rs3745274} being the dominant
pharmacogenetic determinant in the ANRS 12154 study~\citep{Bertrand2012}.

\paragraph{Secondary genetic signals.}
Beyond \texttt{rs3745274}, the consensus heatmap at $\varepsilon=0.01$ highlights a small set of recurring
candidates (Figure~\ref{fig:summary}, Table~\ref{tab:consensus-topk-real}). Of the 15 SNPs in Table 4, all are in \textit{CYP2B6} except for rs2037548 and rs1523129 in \textit{NR1I2}, rs6978925 and rs17149699 in \textit{ABCB1}, and rs4646440 in \textit{CYP3A4}.
Effect-size posteriors for these SNPs are prior-dependent: the Regularized Horseshoe and $\ell_1$-ball allow broader
posteriors and heavier tails, whereas the Hierarchical Lasso and R2--D2 concentrate more mass near zero. This
heterogeneity reflects differences in regularization geometry rather than instability of the NLME component, as core
PK parameters remain stable across priors (Figure~\ref{fig:kde_real}).
Several secondary candidates show moderate correlation with \texttt{rs3745274} ($|\rho|\approx 0.4$--$0.46$) and
moderate allele frequencies, suggesting partial linkage structure rather than fully independent signals.
A notable exception is \texttt{rs1523129}, which is ranked $\approx 2-3^{\mathrm{th}}$ by all continuous priors but is strongly
down-ranked by Spike-and-Slab ($86^{\mathrm{th}}$), yielding a poor average-rank consensus despite appearing among the top
candidates for four of the five baselines (Table~\ref{tab:consensus-topk-real}).
Its very low frequency ($\mathrm{MAF}=0.004$) implies that it is carried by at most one individual; in this setting,
a rare SNP can behave like a subject-identifier covariate and absorb idiosyncratic deviations.
Continuous shrinkage priors, especially heavy-tailed ones, can accommodate such localized likelihood gains by leaving
non-negligible posterior mass away from zero, whereas Spike-and-Slab imposes an explicit inclusion barrier through
its global sparsity parameter and therefore ranks ultra-rare variants more conservatively.

\paragraph{Comparison with univariate association analysis.}
The top consensus SNPs largely overlap with variants reported in the original univariate ANRS 12154 analysis
(Supplementary Table~S1). In particular, the highly ranked CYP2B6 variants (including \texttt{rs3745274},
\texttt{rs4803417}, \texttt{rs10418990}, \texttt{rs1987236}, \texttt{rs2279344}, and \texttt{rs4802101}) have
strong univariate associations ($P \approx 0.000$) and are similarly prioritized by the joint Bayesian NLME model.
This concordance provides external validation while emphasizing that our approach performs selection and uncertainty
propagation within a single coherent hierarchical model, rather than relying on marginal association testing.

Taken together, after accounting for between-occasion variability and the dominant \texttt{rs3745274} effect,
evidence for additional polygenic contributors is modest and remains sensitive to the choice of sparsity mechanism.

\section{Discussion}\label{sec:discussion}

In this study, we evaluated Bayesian variable selection for pharmacogenetic covariates within NLME PK models and found it to be a practical and principled approach when coherent uncertainty propagation and post-selection summaries are important.
Across both simulated and real data, the joint NLME formulation produced stable inference for core PK parameters and predictive checks, while the main differences between sparsity priors concerned how weaker genetic signals were shrunk and ranked.
In simulation, all priors achieved low false-discovery rates under $H_0$ and recovered the causal signal under $H_1$, with peak $F_1$ values around $0.8$--$0.85$ and FWER between $0$ and $0.08$ at $\tau=0.4$.
Among them, Spike-and-Slab is especially attractive when direct inclusion summaries are desired, because it provides analytical PIPs, whereas among priors requiring proxy PIPs the $\ell_1$-ball combined strong recovery with the most stable behavior across $\varepsilon$.
In the ANRS 12154 analysis, PK posteriors and predictive diagnostics remained consistent across priors, whereas SNP rankings beyond \texttt{rs3745274} showed only partial agreement and substantial prior sensitivity.

Posterior inclusion probabilities provide an attractive graded measure of evidence that avoids repeated model fitting and reports uncertainty about inclusion directly.
In our comparison, Spike-and-Slab yields analytical PIPs via Rao--Blackwellization, whereas priors without explicit inclusion indicators require the proxy $\mathbb{P}(|\beta_j|>\varepsilon\mid\mathcal D)$ and therefore an additional tolerance parameter $\varepsilon$.
These Bayesian quantities should not be conflated with frequentist $p$-values: they quantify posterior belief under the joint model and prior, and can be used together with Bayesian error-control criteria or sensitivity analyses across $(\varepsilon,\tau)$, as recommended in Bayesian variable-selection and shrinkage literature~\citep{barbieripip,stephenspip,finemappip,Piironen_2017}.
Practically, Spike-and-Slab also proved more conservative for the ultra-rare variant \texttt{rs1523129}, while the $\ell_1$-ball emerged as the most robust proxy-PIP alternative because its exact-zero projection structure made results comparatively stable across $\varepsilon$.
Reporting both effect-size posteriors and inclusion summaries is therefore essential, since similar rankings can still correspond to different shrinkage of borderline effects.

\subsection{Limitations}

Our NUTS-based implementation targets settings where the SNP dimension is moderate (candidate panels, focused regions, or fine-mapping windows).
Scaling fully Bayesian NLME variable selection to genome-wide applications with $p\approx 10^5$--$10^6$ SNPs is not currently practical with generic HMC, due to the cost of repeatedly updating a high-dimensional coefficient vector and associated local scales.
Addressing this limitation will likely require approximate or specialized inference.
Variational Bayes is a natural direction~\citep{blei2017variational}: mean-field and structured variational approximations for global--local shrinkage priors, as well as amortized or hybrid variational--MCMC strategies, can offer orders-of-magnitude speedups while retaining useful uncertainty summaries in large-$p$ regimes~\citep{komodromos2025group}.
On the MCMC side, algorithmic advances that reduce dependence on $p$ are also promising; for example, sublinear-in-$p$ MCMC schemes for Bayesian variable selection aim to make million-dimensional settings feasible by avoiding full updates of all coefficients at every iteration~ \citep{pmlr-v206-jankowiak23a}.
Finally, pragmatic two-stage workflows remain relevant in pharmacometrics: fast screening or stability-selection methods can reduce the candidate set before joint Bayesian NLME refinement, at the cost of reintroducing some selection dependence~\citep{naveau2024bayesian,gabaut2025stability}.

A second limitation is that sparsity priors require calibration.
To make priors comparable, we tune hyperparameters using explicit sparsity and effect-size targets.
While this strategy is transparent and yields interpretable prior mass near zero versus in the tails, it typically starts from approximate calibration rules (e.g., orthogonal-design proxies and effective model size surrogates) and therefore benefits from being \emph{validated and, if necessary, corrected} by prior simulation.
In practice, we found it important to complement closed-form heuristics with Monte Carlo checks of the induced prior mass in practically relevant effect-size regimes and of the implied (effective) sparsity level, especially in small-$N$ and correlated-SNP settings, where simple proxies may misrepresent the behavior of the full model.
More automated alternatives include placing additional hyperpriors that learn global sparsity from the data, or tuning against predictive criteria when held-out evaluation is feasible, but these approaches must be used carefully to avoid implicit overfitting or overly adaptive regularization.
Developing robust, application-driven calibration guidelines for pharmacogenetic NLME models remains an open problem.

\subsection{Perspectives}

Several extensions could broaden applicability and improve interpretability.
First, richer genetic structure could be incorporated, such as group- or annotation-weighted shrinkage.
Second, the within-subject variability component could be generalized when richer longitudinal sampling is available,
for instance through more flexible occasion-level hierarchies or covariate-dependent variability.
Third, in line with recent pharmacometric work advocating Bayesian shrinkage as a convenient alternative to stepwise
covariate modeling, future work should further connect selection-oriented summaries (PIPs, credible sets) with
predictive evaluation and decision-making objectives in PK/PD applications~\citep{pourzanjani2026rhs}.
Fourth, our current ``consensus'' summary averages posterior inclusion probability SNP ranks across priors; a more principled alternative is Bayesian
multimodel inference, which averages over modeling choices (including sparsity priors) using weights based on
out-of-sample predictive performance~\cite{linden2025increasing}.
Overall, our results indicate that Bayesian variable selection within a joint NLME framework is a coherent and flexible
tool for pharmacogenetic studies, while underscoring the need for scalable inference, principled calibration, and
robust model-averaging strategies in higher-dimensional regimes.

%\section{Conclusion}
\section{Acknowledgments}
Clinical data were collected and analysed thanks to the Agence Nationale de la Recherche contre le Sida (ANRS) funding and genotyping data obtained thanks to NIH grants AI-077505, AI-054999, and TR-000011. DWH receives support from National Institutes of Health (NIH)/NIAID awards R01 AI077505 and P30 AI110527. 

\bibliographystyle{vancouver}
\bibliography{references}

\newpage

\begin{table}[H]
\centering
\small
\begin{tabular}{lll}
\toprule
\textbf{Prior / setting} & \textbf{Simulation study} & \textbf{Real case study} \\
\midrule

\textbf{Spike-and-Slab} & & \\
$\sigma_{\text{spike}}$ & 0.001 & 0.001 \\
$\sigma_{\text{slab}}$  & 0.3   & 0.2   \\
$\theta_a$              & 2     & 2     \\
$\theta_b$              & 70    & 70    \\
\midrule

\textbf{Hierarchical Lasso} & & \\
$\tau_0$ & 0.35 & 0.5 \\
$\rho$   & 70   & 45  \\
\midrule

\textbf{Regularized Horseshoe} & & \\
$\tau_0$              & 0.01  & 0.025 \\
$\nu_{\text{global}}$ & 1     & 1     \\
$\nu_{\text{local}}$  & 1     & 1     \\
$\sigma_{\text{slab}}$& 0.3   & 0.3   \\
$\nu_{\text{slab}}$   & 1.0   & 1.0   \\
\midrule

\textbf{$\ell_1$-ball} & & \\
$b_\eta$     & 0.30 & 0.12 \\
$\lambda_r$  & 1.6  & 3.0  \\
\midrule

\textbf{R2--D2} & & \\
$a$      & 3.5   & 2   \\
$b$      & 397.0 & 30  \\
$\alpha$ & 7.0   & 4.0 \\
\midrule

\textbf{Shared (data scale)} & & \\
$N$            & 400  & 128 \\
$\omega_{CL}$  & 0.30 & 0.22 \\
\bottomrule
\end{tabular}
\caption{
Hyperparameter settings for each sparsity-inducing prior in the simulation study and in the real case analysis.
Effect-size priors are expressed on the log-clearance scale.
}
\label{tab:prior_hyperparameters}
\end{table}

\begin{table}[H]
  \centering
  \caption{Family-Wise Error Rate (FWER) computed across 100 simulated datasets, using analytical posterior inclusion probabilities for the spike-and-slab model and proxy posterior inclusion probabilities with $\varepsilon=0.01$ for the other priors. A posterior inclusion probability threshold of $0.4$ to call discoveries. The reported 95\% confidence intervals are binomial Wilson intervals.}
  \label{tab:ss-top5-analytical-H0}
\begin{tabular}{lccccc}
\toprule
 & Spike-and-Slab & Hierarchical Lasso & Regularized Horseshoe & L1-ball & R2-D2 \\
\midrule
$\mathrm{FWER}$ & 0.040 & 0.000 & 0.020 & 0.080 & 0.000 \\
$95\%$ CI & [0.016, 0.098] & [0.000, 0.037] & [0.006, 0.070] & [0.041, 0.150] & [0.000, 0.037] \\
\bottomrule
\end{tabular}
  \vspace{0.25em}
  \footnotesize\end{table}

\begin{table}[H]
\centering
\begin{tabular}{r l r r r r r r r r}
\toprule
\# & SNP & \makecell{Spike\text{-}and\\\text{-}Slab} & L1\text{-}ball & \makecell{Hierarchical\\Lasso} & \makecell{Regularized\\Horseshoe} & R2\text{-}D2 & Mean & $\rho$ & MAF \\
\midrule
1 & \texttt{rs3745274} & 1 & 1 & 1 & 1 & 1 & 1.0 & -- & 0.348 \\
2 & \texttt{rs4803417} & 3 & 3 & 3 & 4 & 2 & 3.0 & -0.45 & 0.332 \\
3 & \texttt{rs10418990} & 4 & 4 & 4 & 6 & 4 & 4.4 & -0.44 & 0.328 \\
4 & \texttt{rs1987236} & 5 & 7 & 5 & 8 & 6 & 6.2 & -0.46 & 0.328 \\
5 & \texttt{rs2279344} & 2 & 6 & 7 & 16 & 5 & 7.2 & -0.40 & 0.277 \\
6 & \texttt{rs4802101} & 15 & 5 & 6 & 9 & 7 & 8.4 & -0.44 & 0.324 \\
7 & \texttt{rs2037548} & 14 & 15 & 8 & 5 & 11 & 10.6 & -0.01 & 0.012 \\
8 & \texttt{rs6978925} & 19 & 11 & 13 & 3 & 14 & 12.0 & +0.05 & 0.234 \\
9 & \texttt{rs4802100} & 10 & 10 & 22 & 12 & 13 & 13.4 & -0.21 & 0.113 \\
10 & \texttt{rs17149699} & 20 & 14 & 15 & 13 & 12 & 14.8 & -0.01 & 0.324 \\
11 & \texttt{rs4646440} & 16 & 17 & 18 & 24 & 19 & 18.8 & -0.05 & 0.266 \\
12 & \texttt{rs11671243} & 18 & 16 & 12 & 39 & 10 & 19.0 & -0.39 & 0.285 \\
13 & \texttt{rs1523129} & 86 & 2 & 2 & 2 & 3 & 19.0 & +0.04 & 0.004 \\
14 & \texttt{rs1808682} & 38 & 12 & 14 & 17 & 15 & 19.2 & -0.34 & 0.207 \\
15 & \texttt{rs892216} & 6 & 28 & 29 & 22 & 28 & 22.6 & -0.46 & 0.324 \\
\bottomrule
\end{tabular}
\caption{Top consensus SNPs ranked by average rank across priors.}
\label{tab:consensus-topk-real}
\end{table}

\newpage

\clearpage
\appendix

\renewcommand{\thefigure}{S\arabic{figure}}
\renewcommand{\thesection}{S\arabic{section}}
\renewcommand\thetable{S\arabic{table}}
\renewcommand\theequation{S\arabic{equation}}

\setcounter{figure}{0}
\setcounter{equation}{0}
\setcounter{section}{0}
\setcounter{table}{0}

\begin{figure}[b]
    \centering
    \includegraphics[width=1\linewidth]{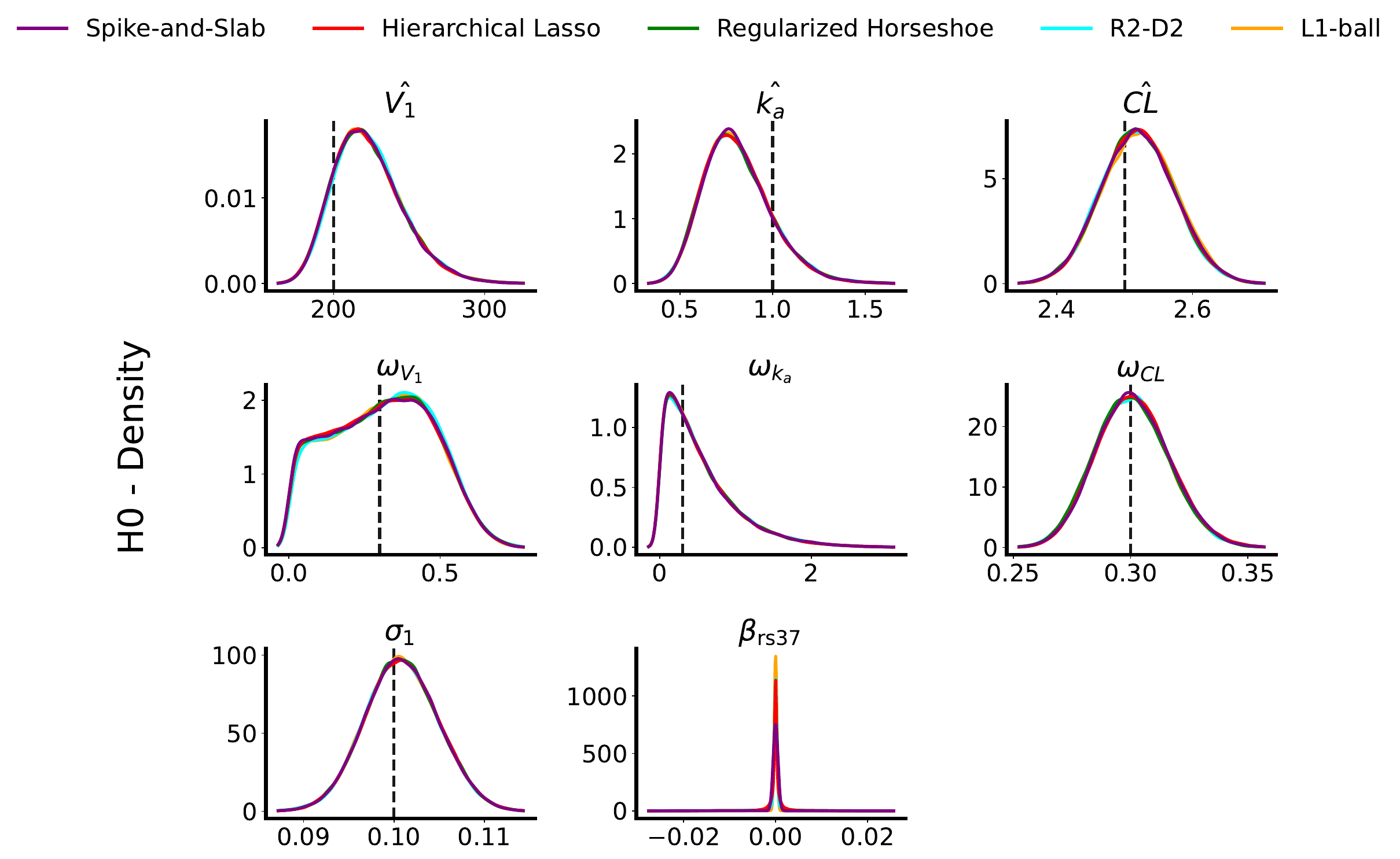}
    \includegraphics[width=1\linewidth]{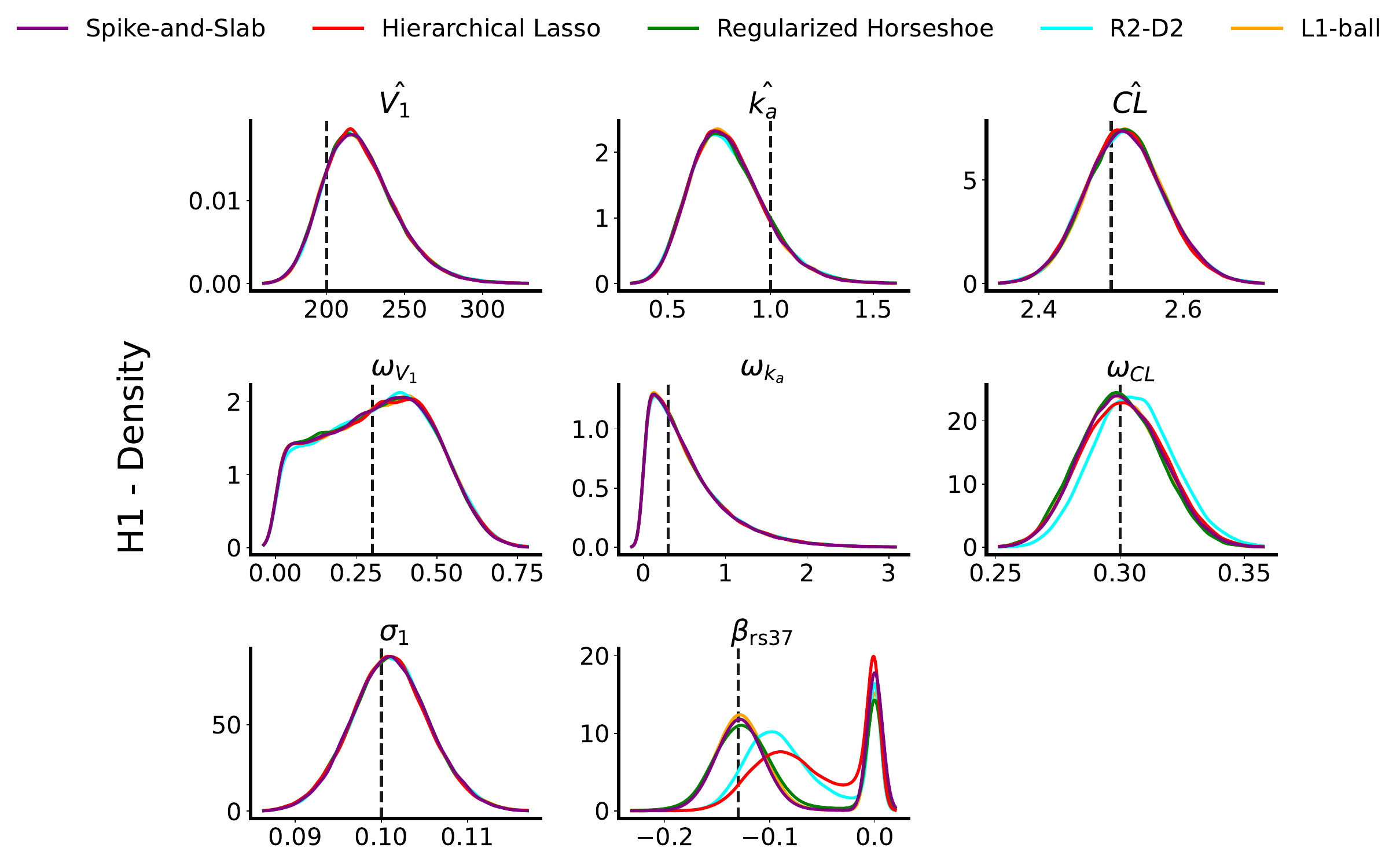}
    \caption{Posterior distributions of pharmacokinetic parameters under $H_0$ (top) and $H_1$ (bottom) across the five sparsity-inducing priors. Shown are the population parameters, their inter-individual variability terms, the residual error, and the SNP effect $\beta_{\text{rs37}}$. Kernel density estimates are based on draws from 5 NUTS chains with 2000 samples each, aggregated over 100 simulated datasets.}
    \label{fig:kde}
\end{figure}

\begin{figure}[b]
    \centering
    \includegraphics[width=1\linewidth]{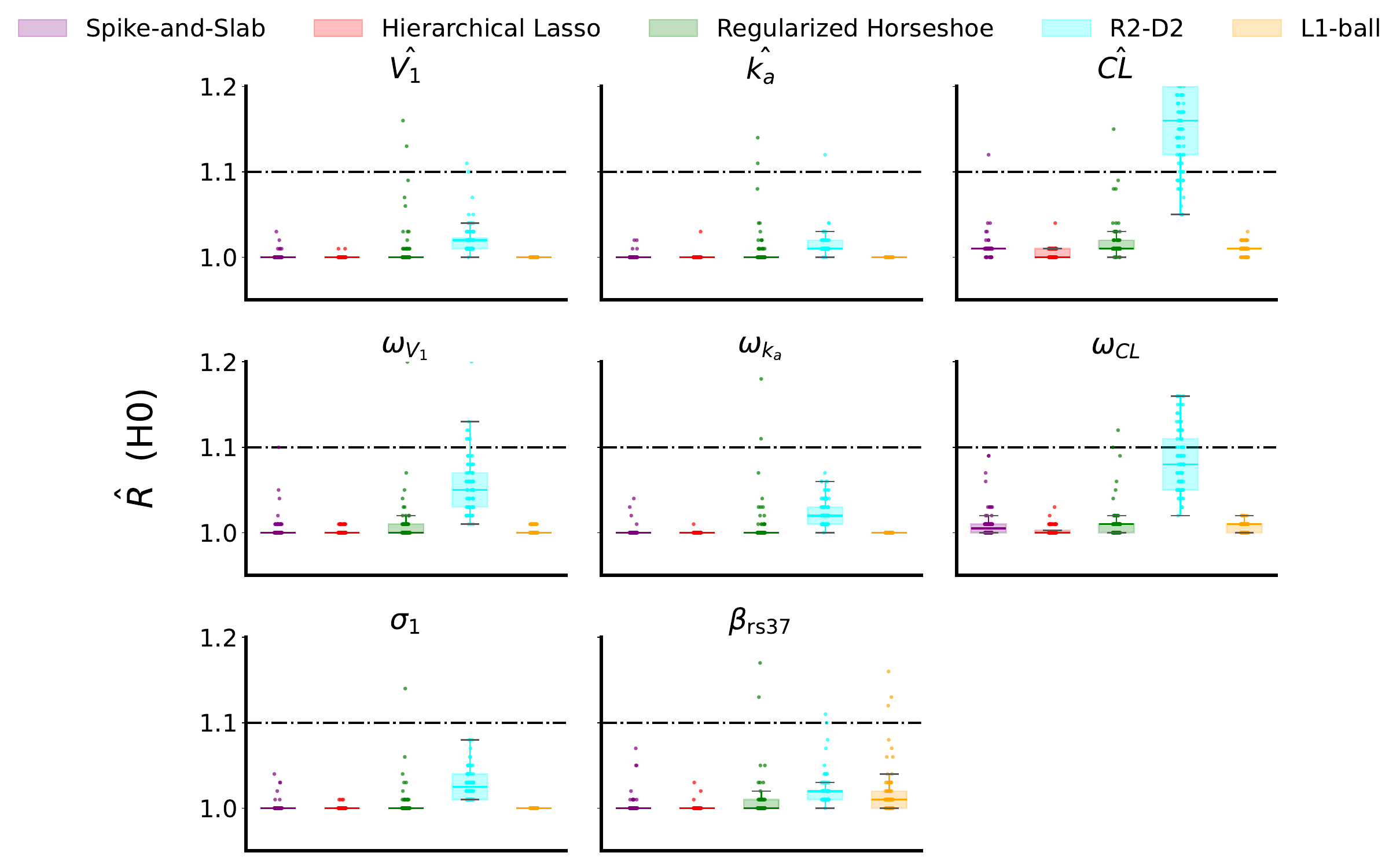}
    \includegraphics[width=1\linewidth]{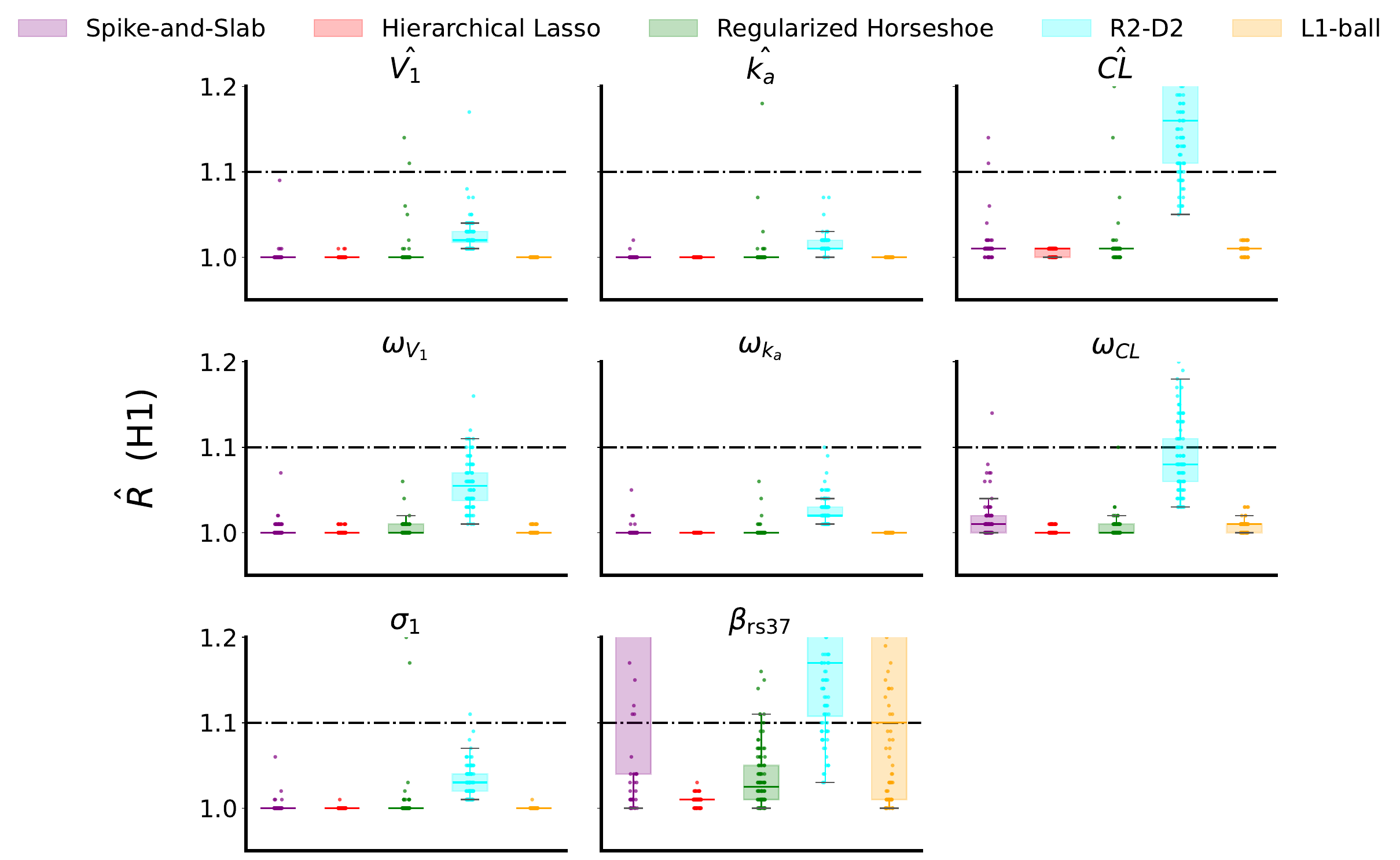}
    \caption{Convergence diagnostics under $H_0$ (top) and $H_1$ (bottom). Distribution of $\hat{R}$ values computed using 5 NUTS chains for the main pharmacokinetic parameters, their inter-individual variability terms, the residual error, and the SNP effect $\beta_{\text{rs37}}$. Results are shown across 100 simulated datasets for each of the five sparsity-inducing priors. The dashed line at $\hat{R}=1.1$ marks a common threshold for acceptable convergence}
    \label{fig:rhat}
\end{figure}

\begin{figure}[b]
    \centering
    \includegraphics[width=1\linewidth]{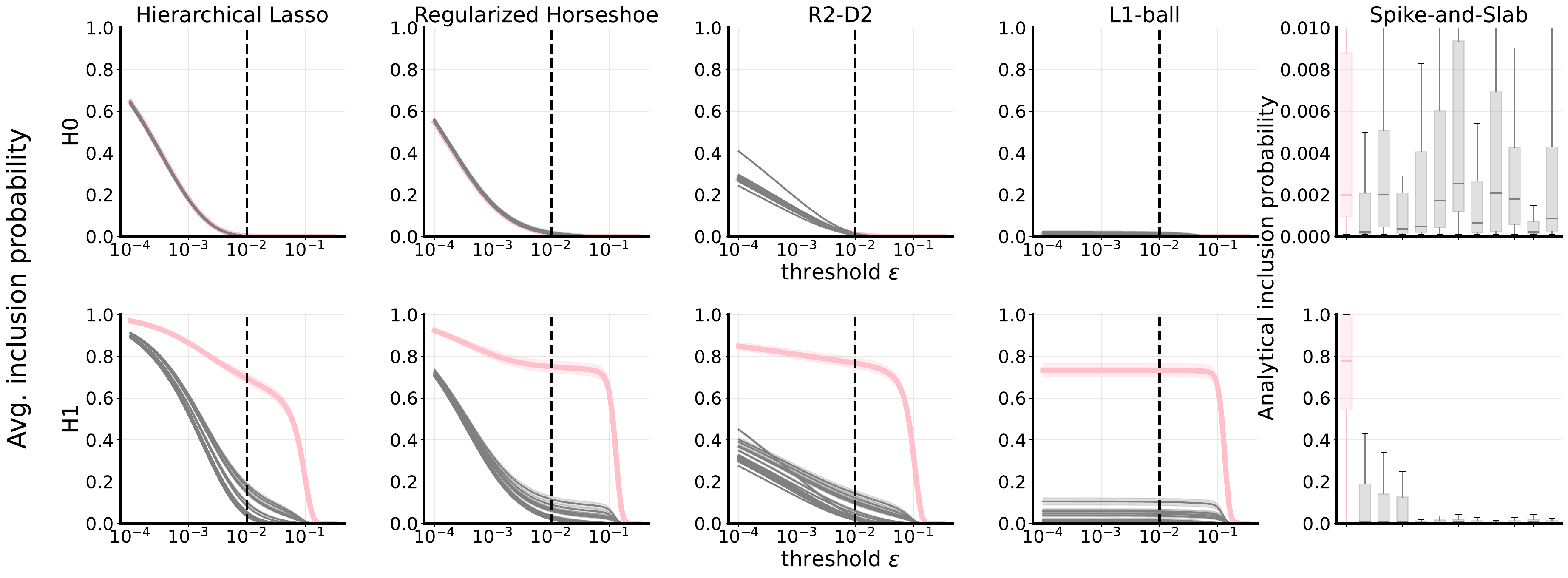}
    \caption{Average posterior inclusion probabilities $\pm$ standard error across 100 simulated datasets under $H_0$ (top row) and $H_1$ (bottom row). 
    For the continuous shrinkage priors (Hierarchical Lasso, Regularized Horseshoe, R2-D2, L1-ball), posterior inclusion probabilities are defined as $p(\lvert \beta_j\rvert > \varepsilon)$ for varying thresholds $\varepsilon$ (x-axis, log scale). The vertical dashed line marks $\varepsilon=0.01$, the default used in the main analyses. For the spike-and-slab prior, analytical posterior inclusion probabilities are available directly and therefore shown without dependence on $\varepsilon$, and only the 10 SNPs with the highest posterior probability of inclusion are shown.}
    \label{fig:inclprob}
\end{figure}

\newpage

\appendix

\section{Additional details for the sparsity-enforcing priors}\label{sec:detailspriors}

\subsection{Rationale for the regularized horseshoe}

Relative to the classical horseshoe, the regularized form modifies the effective shrinkage factor by combining the local scale with a slab term $\tilde{\lambda}_j^2 = \frac{c^2 \lambda_j^2}{c^2 + \tau^2 \lambda_j^2}$. The parameter $c$ sets the magnitude at which the slab begins to act. When $(\tau \lambda_j)^2 \ll c^2$, the behavior matches the classical horseshoe with heavy tails, whereas for $(\tau \lambda_j)^2 \gg c^2$, shrinkage is damped and coefficients are prevented from growing without bound. In this way, $c$ controls the transition between heavy-tailed behavior and regularization, capping the influence of very large $\lambda_j$, while still allowing moderately large effects to remain plausible. We refer to~\citep{Piironen_2017} for an in-depth analysis.

The resulting prior has two key properties: polynomial concentration near zero (stronger than the horseshoe) and polynomially heavy tails (heavier than Cauchy when $b < 1/2$), yielding near-minimax contraction rates comparable to spike-and-slab but with continuous shrinkage. This balance of sharp sparsity and robust tail behavior makes R2-D2 one of the most powerful shrinkage priors in high-dimensional regression.

\section{Further details on prior calibration}\label{app:calibration}

\subsection{Tail probabilities}\label{app:ssdetails}

We recall the computation of two-sided tail probabilities in the case of a Gaussian distribution,
e.g., for computing Equation~\ref{eq:tailss} in the Spike-and-Slab model.

Under the scale-invariant parameterization,
\[
\beta_j = \omega_h \tilde{\beta}_j,
\]
and the slab prior is written directly on $\beta_j$ as
\[
\beta_j \mid z_j=1 \sim \mathcal N\!\left(0,\ \omega_h^2 \sigma_{\text{slab}}^2\right).
\]
Equivalently, $\tilde{\beta}_j \mid z_j=1 \sim \mathcal N(0,\sigma_{\text{slab}}^2)$.

For any $a>0$,
\begin{align*}
p\big(|\beta_j|>a \mid z_j=1\big)
&= p(\beta_j>a \mid z_j=1) + p(\beta_j<-a \mid z_j=1) \\
&= 2p(\beta_j>a \mid z_j=1) \quad\text{(symmetry)}\\
&= 2p\Big(\frac{\beta_j}{\omega_h\sigma_{\text{slab}}} 
> \frac{a}{\omega_h\sigma_{\text{slab}}} \ \Big|\ z_j=1\Big)\\
&= 2\Big(1-\Phi\big(a/(\omega_h\sigma_{\text{slab}})\big)\Big),
\end{align*}
where $\Phi$ denotes the standard Normal CDF.

With $a=0.13$ and $\sigma_{\text{slab}}=0.1$, this becomes
\[
p(|\beta_j|>0.13\mid z_j=1)
= 2\Big(1-\Phi\big(0.13/(\omega_h\times0.1)\big)\Big).
\]

Similarly, for the spike
\[
\beta_j\mid z_j=0 \sim 
\mathcal N\!\left(0,\ \omega_h^2\sigma_{\text{spike}}^2\right),
\]
we obtain
\[
p(|\beta_j|>a \mid z_j=0)
= 2\Big(1-\Phi\big(a/(\omega_h\sigma_{\text{spike}})\big)\Big).
\]

For example, with $a=0.01$ and $\sigma_{\text{spike}}=0.0005$,
\[
p(|\beta_j|>0.01 \mid z_j=0)
= 2\Big(1-\Phi\big(0.01/(\omega_h\times0.0005)\big)\Big),
\]
which is effectively zero for typical values of $\omega_h$.

% In other words, $\tau$ governs the overall sparsity level, while the local scales $\lambda_j$ let individual coefficients adapt by escaping shrinkage when strongly supported by the data.

\subsection{Derivation of the shrinkage factor $\kappa_j$ for global-local shrinkage priors}
\label{app:shrinkagefactor}

Here, we justify Equations~\ref{eq:coordlike} and~\ref{eq:kappa} introduced in the main text.
We start from the standard linear-Gaussian proxy model with centered columns:
\begin{equation}
\y = \X\bbeta + \beeta,\qquad 
\beeta\sim\mathcal N(\mathbf 0,\ \omega_{h}^2 \I_N).
\end{equation}

The (unnormalized) likelihood is
\begin{equation}
\mathcal{L}(\bbeta)
\propto
\exp\!\left(-\frac{1}{2\omega_{h}^2}
\|\mathbf y - \X\bbeta\|_2^2\right).
\end{equation}

Now isolate coordinate $j$. Write
\begin{equation}
\X\bbeta 
= \sum_{k=1}^p \mathbf x_k \beta_k 
= \mathbf x_j \beta_j 
+ \sum_{k\neq j}\mathbf x_k \beta_k,
\end{equation}
and define the residual after removing all but $j$:
\begin{equation}
\mathbf r_j(\beta_{-j})
:= 
\y - \sum_{k\neq j}\mathbf x_k \beta_k.
\end{equation}

Then
\begin{equation}
\| \y - \X\bbeta\|_2^2
=
\|\mathbf r_j - \mathbf x_j\beta_j\|_2^2
=
\|\mathbf r_j\|_2^2 
- 2\beta_j\mathbf x_j^\top \mathbf r_j 
+ \beta_j^2\|\mathbf x_j\|_2^2.
\end{equation}

Under the orthogonal-design assumption,
$\X^\top \X = \mathrm{diag}(\|\mathbf x_1\|_2^2,\ldots,\|\mathbf x_p\|_2^2)$
and, with standardization, $\|\mathbf x_j\|_2^2 = N$.
Orthogonality gives
\begin{equation}
\mathbf x_j^\top \mathbf r_j
=
\mathbf x_j^\top
\!\left(
\mathbf y - \sum_{k\neq j}\mathbf x_k \beta_k
\right)
=
\mathbf x_j^\top \mathbf y,
\end{equation}
since $\mathbf x_j^\top \mathbf x_k = 0$ for $k\neq j$.

Plugging back,
\begin{equation}
\|\y - \X\bbeta\|_2^2
=
\|\mathbf r_j\|_2^2 
- 2\beta_j\mathbf x_j^\top \mathbf y 
+ \beta_j^2 N.
\end{equation}

Complete the square in $\beta_j$. Define the least-squares coefficient
\begin{equation}
\hat\beta_j 
:= 
\frac{\mathbf x_j^\top \mathbf y}{\|\mathbf x_j\|_2^2}
=
\frac{\mathbf x_j^\top \mathbf y}{N}.
\end{equation}

Then
\begin{equation}
- \frac{1}{2\omega_{h}^2}
\Big(
\beta_j^2 N - 2\beta_j\mathbf x_j^\top\mathbf y
\Big)
=
-\frac{N}{2\omega_{h}^2}
(\beta_j - \hat\beta_j)^2
+ \text{const}.
\end{equation}

Therefore, holding other coordinates implicit, the likelihood as a function of $\beta_j$ is
\begin{equation}
\mathcal{L}_j(\beta_j)
\propto
\exp\!\left(
-\frac{N}{2\omega_{h}^2}
(\beta_j-\hat\beta_j)^2
\right),
\end{equation}
i.e., Gaussian with mean $\hat\beta_j$ and variance $\omega_{h}^2/N$.

\vspace{0.5em}

Next, we perform the conjugate Gaussian update.
Recall the local weights defined in the main text:
\begin{equation}
w_j =
\begin{cases}
\lambda_j^{2}          & \text{Hierarchical Lasso (Eq.~\ref{eq:betahl})},\\[4pt]
\tilde{\lambda}_j^{2}  & \text{Regularized Horseshoe (Eq.~\ref{eq:betahs})},\\[4pt]
\lambda_j              & \text{R2--D2 (Eq.~\ref{eq:betar2})}.
\end{cases}
\label{eq:localweightapp}
\end{equation}

Under the scale-invariant parameterization,
\begin{equation}
\beta_j \mid \tau,w_j 
\sim 
\mathcal N\!\big(0,\ \omega_{h}^2\tau^2w_j\big).
\end{equation}

Write prior variance $v_0=\omega_{h}^2\tau^2 w_j$
and likelihood variance $v_1=\omega_{h}^2/N$.
The Gaussian--Gaussian posterior mean is the precision-weighted average:
\begin{equation}
\mathbb E[\beta_j\mid \mathbf y,\tau,w_j]
=
\frac{\hat\beta_j/v_1}{1/v_0+1/v_1}.
\end{equation}

Substituting $v_0$ and $v_1$,
\begin{equation}
\mathbb E[\beta_j\mid \mathbf y,\tau,w_j]
=
\frac{\tfrac{N}{\omega_{h}^2}\hat\beta_j}
     {\tfrac{1}{\omega_{h}^2\tau^2 w_j}
      +\tfrac{N}{\omega_{h}^2}}
=
\frac{N\hat\beta_j}
     {N + (\tau^2 w_j)^{-1}}.
\end{equation}

Define the coordinate-wise shrinkage factor
\begin{equation}
\kappa_j 
:= 
\frac{1}{1+N\tau^2 w_j}.
\end{equation}

Then
\begin{equation}
\frac{N}{N + (\tau^2 w_j)^{-1}}
=
1-\kappa_j,
\end{equation}
and therefore
\begin{equation}
\mathbb E[\beta_j\mid \mathbf y,\tau,w_j]
=
(1-\kappa_j)\hat\beta_j,
\qquad
\kappa_j=\frac{1}{1+N\tau^2 w_j}.
\label{eq:shrinkage}
\end{equation}

Under orthogonality, each coordinate decouples, and the posterior mean takes the ridge-type form above.
Substituting the prior-specific definition of $w_j$ (Eq.~\ref{eq:localweightapp}) yields the shrinkage factor for each global--local prior.

\subsection{Choice of hyperparameters}\label{app:hpval}

We now link $\mathbb{E}[m_{\mathrm{eff}}] =\sum_{j=1}^p \mathbb{E}(1-\kappa_j)$
to prior hyperparameters using
Equation~\ref{eq:shrinkage} with $\kappa_j = (1 + N\tau^2 w_j)^{-1}.$

\paragraph{Hierarchical Lasso (Equation~\ref{eq:betahl}).}
Here, $w_j=\lambda_j^2$ in Equation~\ref{eq:shrinkage}, with $\lambda_j\sim \mathrm{Exp}(r)$, yielding
$\kappa_j = (1 + N\tau^2 \lambda_j^2)^{-1}$.
To calibrate $\tau_0$, we target
$\mathbb{E}(1-\kappa_j)\approx \tilde m/p$
using the Jensen plug-in
\begin{equation}
\mathbb{E}[\kappa_j]
\approx
\frac{1}{1 + N\tau^2\mathbb{E}[\lambda_j^2]},
\qquad
\mathbb{E}[\lambda_j^2]=\frac{2}{r^2}.
\end{equation}
This gives
\begin{equation}
\frac{\tilde m}{p}
\approx
\frac{N\tau^2\frac{2}{r^2}}
     {1 + N\tau^2\frac{2}{r^2}}
\quad\Longrightarrow\quad
\tau_0
\approx
\frac{r}{\sqrt{2N}}
\sqrt{\frac{\tilde m}{p-\tilde m}}.
\end{equation}
\emph{Remark.}
Since $x\mapsto(1+ax)^{-1}$ is convex for $x\ge0$,
$\mathbb{E}[\kappa_j]\ge (1+N\tau^2\mathbb{E}[\lambda_j^2])^{-1}$;
thus the plug-in slightly overestimates
$\mathbb{E}(1-\kappa_j)$,
so $\tau_0$ is mildly conservative.

A crude tail proxy treats
$\beta_j$ as
$\mathcal{N}(0,\omega_h^2\tau_0^2\mathbb{E}[\lambda_j^2])$,
giving

\begin{equation}
p(|\beta_j|>a)
\approx
2\left(
1-\Phi\!\left(
\frac{a}{\omega_h\tau_0\sqrt{2}/r}
\right)
\right).
\end{equation}

\paragraph{Regularized Horseshoe (Equation~\ref{eq:betahs}).}
Parts of this derivation directly follow from~\cite{Piironen_2017}.
Local scales $\lambda_j>0$, global $\tau>0$, and a Gaussian slab (scale $c>0$) yield the regularized horseshoe (RHS) local factor
\begin{equation}
\tilde\lambda_j^2 = \frac{c^2 \lambda_j^2}{c^2 + \tau^2 \lambda_j^2},\qquad
\kappa_j = \frac{1}{1 + N\tau^2\tilde\lambda_j^{2}}.
\end{equation}
\cite{Piironen_2017} show that this is well approximated by an affine transform of the classical horseshoe (HS) shrinkage:
\begin{equation}
    \tilde{\kappa}_j \approx (1-b)\kappa_j + b \quad \text{with } b = \frac{1}{1+Nc^2}.
\end{equation}
Intuitively, the slab shifts the shrinkage profile upward from $(0, 1)$ to $(b_j,1)$, preventing shrinkage for very large signals. 
Then, the prior effective size rescales as
\begin{equation}
 \tilde{m}_{\text{eff, RHS}} \approx (1-b)m_{\text{eff, HS}},
 \label{eq:meff}
\end{equation}
meaning that the regularized horseshoe implies a lower model complexity than the classical horseshoe.
\emph{Via} ~\cite{Piironen_2017}[Equation 3.12], targeting a prior effective size $\tilde m$ for the classical horseshoe yields a global scale variance $\tau_0$ such that
\begin{equation}
\tau_{0,\text{HS}} \approx \frac{1}{\sqrt{N}}\ \frac{\tilde m}{p-\tilde m}.
\end{equation}
And using Equation~\ref{eq:meff}, targeting an effective model size $\tilde{m}$ in the regularized horseshoe is equivalent to targeting $\tilde{m}/(1-b)$ under the classical horseshoe. Thus:
\begin{equation}
\tau_{0,\text{RHS}} \approx \frac{1}{\sqrt{N}}\frac{\tilde{m}}{(1-b)p - \tilde{m}}.
\end{equation}
Since $b>0$, the denominator is smaller than $p-\tilde{m}$, making $\tau_{0,\text{RHS}}$ slightly larger than $\tau_{0,\text{HS}}$, compensating for the slab's extra regularization.

The slab $c$ controls magnitude:
\begin{equation}
p(|\beta_j|>a \mid \text{slab})
=
2\left(
1-\Phi\left(
\frac{a}{\omega_h c}
\right)
\right).
\end{equation}
In practice:
(i) set $\tau=\tau_0$,
(ii) tune $c$ by short prior simulation,
(iii) recheck $\mathbb{E}[m_{\mathrm{eff}}]$.

\paragraph{R2--D2  (Equation~\ref{eq:betar2}).}
With $w_j = \lambda_j$, the shrinkage factor is
\begin{equation}
\kappa_j=\frac{1}{1+N\tau^2\lambda_j},
\end{equation}
By Jensen’s inequality applied to the concave map $x\mapsto \frac{N x}{1+N x}$ on $x\ge 0$,
\begin{equation}
\mathbb{E}(1-\kappa_j)
=\mathbb{E}\!\left[\frac{N\tau^2\lambda_j}{1+N\tau^2\lambda_j}\right]
\ \le\
\frac{N\mathbb{E}[\tau^2\lambda_j]}{1+N\mathbb{E}[\tau^2\lambda_j]}.
\end{equation}
Since $\tau^2$ and $\lambda_j$ are independent in the R2--D2 prior, and $\mathbb{E}[\lambda_j]=1/p$ for
$(\lambda_1,\ldots,\lambda_p)\sim\mathrm{Dirichlet}(\alpha,\ldots,\alpha)$,
\begin{equation}
\mathbb{E}[\tau^2\lambda_j]=\frac{1}{p}\mathbb{E}[\tau^2].
\end{equation}
Finally, using the plug-in for the convex map $u\mapsto \frac{u}{1-u}$,
\begin{equation}
\mathbb{E}[\tau^2]=\mathbb{E}\!\left[\frac{R^2}{1-R^2}\right]\ \approx\ \frac{\mathbb{E}[R^2]}{1-\mathbb{E}[R^2]},
\end{equation}
we obtain
\begin{equation}
\mathbb{E}(1-\kappa_j)\ \approx\
\frac{\displaystyle \frac{N}{p}\frac{\mathbb{E}[R^2]}{1-\mathbb{E}[R^2]}}
{\displaystyle 1+\frac{N}{p}\frac{\mathbb{E}[R^2]}{1-\mathbb{E}[R^2]}}.
\end{equation}
To target a prior effective size $\tilde m$, solve $\mathbb{E}(1-\kappa_j)\approx \tilde m/p$ for the mean of $R^2$:
\begin{equation}
\mathbb{E}[R^2]
\approx
\frac{\frac{p}{N}\frac{\tilde m}{p-\tilde m}}
     {1+\frac{p}{N}\frac{\tilde m}{p-\tilde m}}.
\end{equation}
Choose $(a,b)$ such that
$a/(a+b)$ equals this value.

\emph{Remark.} The first step is an upper bound (Jensen on the concave $x\mapsto \tfrac{N x}{1+N x}$);
the second plug-in (convex $u\mapsto \tfrac{u}{1-u}$) slightly underestimates $\mathbb{E}[\tau^2]$, so the mapping is mildly conservative.
\subsection{Soft--thresholding and memorylessness for $\ell_1$-ball calibration}

Let $\delta\ge 0$ be the common soft--threshold determined by the projection. Define the active magnitude
\begin{equation}
V_j := (U_j-\delta)_+ .
\end{equation}
On the event ``active'' ($U_j>\delta$), we have $V_j = U_j-\delta$. For any fixed $\delta\ge 0$ and $t\ge 0$,
\begin{equation}
p(V_j \le t \mid U_j>\delta,\ \delta)
= p(U_j-\delta \le t \mid U_j>\delta,\ \delta)
= p(U_j \le t+\delta \mid U_j>\delta).
\end{equation}
Using the memoryless property of the exponential,
\begin{equation}
p(U_j \le t+\delta \mid U_j>\delta)
= 1 - \frac{p(U_j>t+\delta)}{p(U_j>\delta)}
= 1 - \frac{e^{-\lambda(t+\delta)}}{e^{-\lambda\delta}}
= 1 - e^{-\lambda t}.
\end{equation}
Thus the conditional law is exponential with the \emph{same} rate:
\begin{equation}
V_j \big| (U_j>\delta,\ \delta) \ \sim\ \mathrm{Exp}(\lambda)\quad\text{on }[0,\infty).
\end{equation}
Although $\delta$ is random (it depends on the whole vector $\boldsymbol{\xi}$ and $r$),
the above holds \emph{for every realized value of $\delta$}. Therefore, marginalizing over $\delta$ preserves the exponential form:
\[
V_j \big| (\text{active}) \ \sim\ \mathrm{Exp}(\lambda=1/b_\xi).
\]

Under our scale-invariant parameterization $\beta_j=\omega_h\tilde\beta_j$ with
$\tilde{\xi}_j\sim\mathrm{Laplace}(0,b_\xi)$ and $\tilde r\sim\mathrm{Exp}(\lambda_r)$,
the active magnitudes satisfy $|\tilde\beta_j|\mid(\tilde\beta_j\neq 0)\sim \mathrm{Exp}(1/b_\xi)$.
Equivalently, on the original scale, $|\beta_j|\mid(\beta_j\neq 0)$ has mean $\omega_h b_\xi$ and tail
\[
p(|\beta_j|>a \mid \beta_j\neq 0)=\exp\!\left(-\frac{a}{\omega_h b_\xi}\right),
\]
which we use to set $b_\xi$ (e.g., with $b_\xi=0.10$ the tail at $a=0.13$ is $\exp(-0.13/(\omega_h\times0.10))$).

Finally, because the projection enforces $\sum_{j:\tilde\beta_j\neq 0}|\tilde\beta_j|=\tilde r$ and each active $|\tilde\beta_j|$ has mean $b_\xi$,
a first--order approximation gives
\[
\mathbb{E}[K]\approx \frac{\mathbb{E}[\tilde r]}{b_\xi} = \frac{1}{\lambda_r b_\xi},
\]
which we use to hit a target effective size (e.g., $\tilde m\approx 5$).

\begin{figure}[b]
\centering

% --- Top-left: PDF plot ---
\begin{subfigure}{0.32\textwidth}
  \centering
  \includegraphics[width=\linewidth]{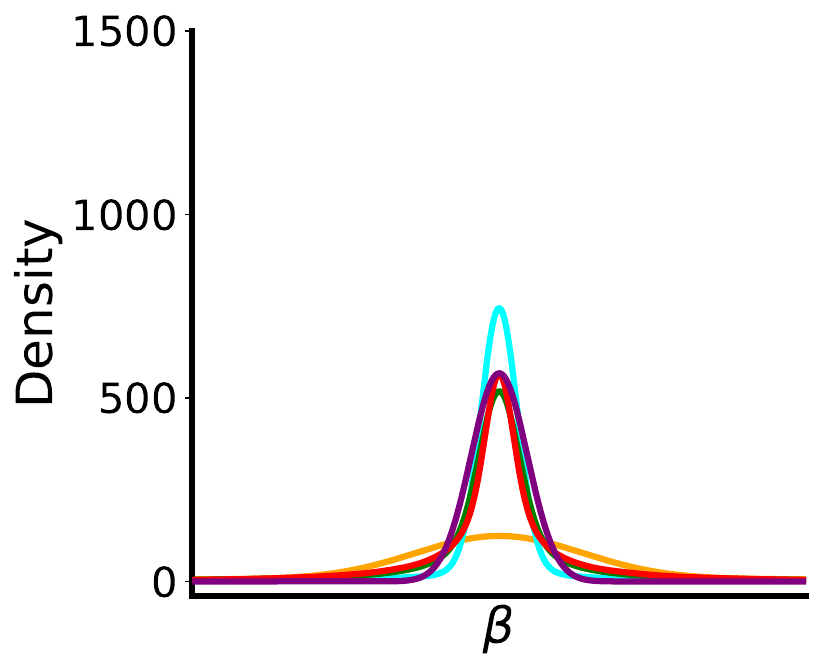} % <- replace filename
  \caption{}
  \label{fig:priors-superposedreal}
\end{subfigure}
\hfill
% --- Top-right: Table ---
\begin{subfigure}{0.67\textwidth}
  \centering
  \renewcommand{\arraystretch}{1.3}   % vertical padding
  \setlength{\tabcolsep}{2pt}         % horizontal padding
  {\small
\begin{tabular}{lccccc}
\toprule
 & Spike-and-Slab & L1-ball & Hierarchical Lasso & Regularized Horseshoe & R2-D2 \\
\midrule
$\mathbb{P}(|\beta|<10^{-3})$ & 0.973 & 0.612 & 0.650 & 0.702 & 0.861 \\
$\mathbb{P}(|\beta|<10^{-2})$ & 0.977 & 0.725 & 0.977 & 0.964 & 0.967 \\
$\mathbb{P}(10^{-2}<|\beta|<10^{-1})$ & 0.022 & 0.266 & 0.023 & 0.036 & 0.033 \\
$\mathbb{P}(|\beta|>10^{-2})$ & 0.023 & 0.275 & 0.023 & 0.036 & 0.033 \\
$\mathbb{P}(|\beta|>10^{-1})$ & 0.001 & 0.009 & 0.000 & 0.000 & 0.000 \\
$\mathbb{P}(|\beta|>10^{-1}\mid \beta\neq 0)$ & \textcolor{red}{\xmark} & 0.022 & \textcolor{red}{\xmark} & \textcolor{red}{\xmark} & \textcolor{red}{\xmark} \\
\bottomrule
\end{tabular}}
  \caption{}
  \label{fig:prior-table-statisticsreal}
\end{subfigure}

\vspace{0.5em}
\begin{subfigure}{.5\textwidth}
\centering
  \includegraphics[width=1.5\linewidth]{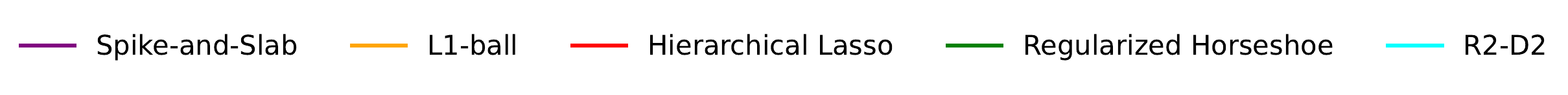}
\end{subfigure}
\vspace{0.75em}
% --- Bottom: single wide PDF plot ---
\begin{subfigure}{\textwidth}
  \centering
  \includegraphics[width=\linewidth]{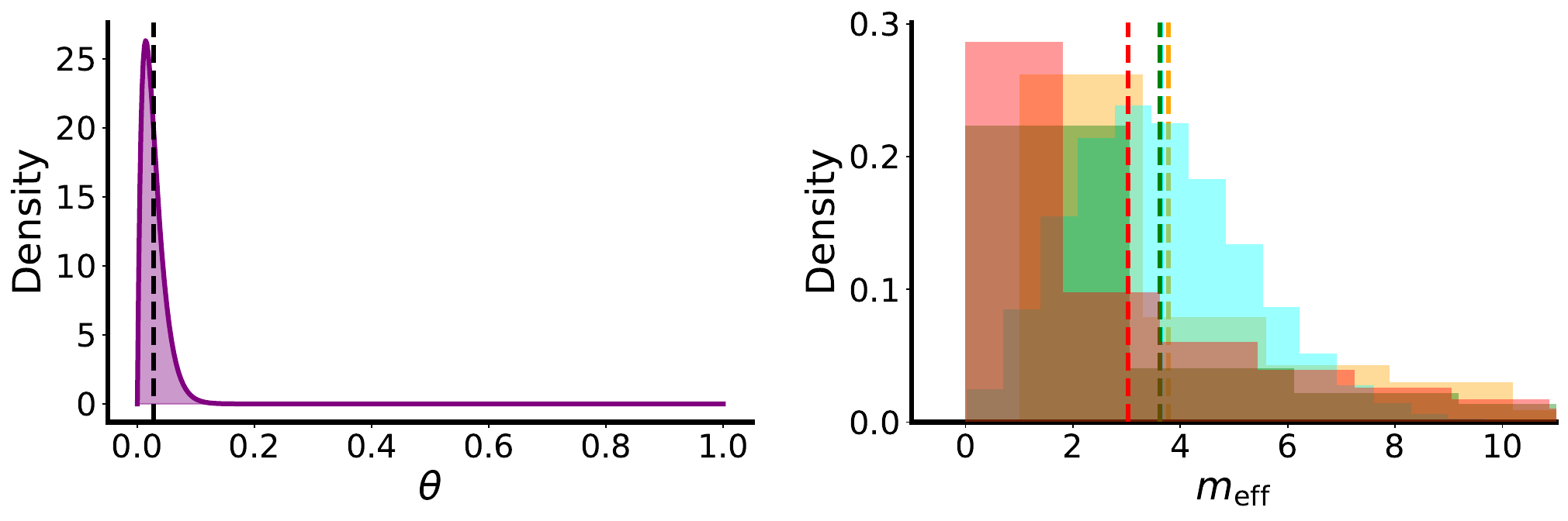}
  \caption{}
  \label{fig:priorchecksreal}
\end{subfigure}

\caption{
Prior checks for real data.
(a) Marginal prior distributions for a given SNP effect size $\beta_j$ under the five sparsity-enforcing priors. (b) Prior probabilities of different effect size regimes (with $10^{-3}$ precision), summarizing the mass near zero and in the tails. (c) Left: Beta$(1,30)$ distribution for the global inclusion probability $\theta$ in the spike-and-slab prior. Right: distribution of the effective number of nonzero coefficients $m_{\text{eff}}$ in a SNP effect size vector $\boldsymbol{\beta}$. Kernel density estimates, histograms, and summary statistics are based on 200,000 prior draws.}
\label{fig:fig1real}
\end{figure}

\section{Pseudo-codes for sparsity-enforcing priors}\label{sec:pseudocodes}

All implementations are given in \texttt{NumPyro}~\citep{bingham2019pyro, phan2019composable}, a probabilistic programming library built on top of \texttt{JAX}~\cite{jax2018github}, which enables automatic differentiation and GPU acceleration. We provide pseudo-code listings for each prior considered in the main text, to clarify how they are realized in practice.

Of note, for the R2-D2 prior, the allocation of the total variance budget across predictors is governed by a symmetric Dirichlet distribution. In our implementation, this allocation is generated using an exact stick-breaking construction~\citep{blei_stickbreaking}: at each step, a proportion of the remaining “stick” (total variance) is broken off according to a Beta distribution and assigned to a predictor. Iterating this procedure yields a set of weights $\{\phi_j\}_{j=1}^p$ that sum to one:
\begin{equation}
\phi_1 = v_1,\quad \phi_k = v_k \prod_{\ell<k}(1-v_\ell),\quad \phi_p = \prod_{\ell=1}^{p-1}(1-v_\ell).
\end{equation}
While mathematically equivalent to drawing directly from a Dirichlet distribution, the stick-breaking construction is convenient in probabilistic programming because it decomposes the simplex constraint into a sequence of independent Beta draws, making the implementation more stable.

\begin{lstlisting}[language=Python, caption=NumPyro pseudo-code for the spike--and--slab prior,label={lst:spike_slab_numpyro}]
import numpyro
import numpyro.distributions as dist
import jax.numpy as jnp
from jax.nn import logsumexp

def spike_slab_prior(nSNP, spike_sd, slab_sd, omegaCL, gamma_a, gamma_b):

    # Inclusion probability
    gamma = numpyro.sample("theta", dist.Beta(gamma_a, gamma_b))
    
    # SNP-wise inclusion indicators z_j are marginalized in parallel
    with numpyro.plate("snps", nSNP):
        z = numpyro.sample("z", dist.Bernoulli(probs=gamma),
                           infer={"enumerate": "parallel"})
        scale = omegaCL * (spike_sd * (1 - z) + slab_sd * z)
        beta = numpyro.sample("beta", dist.Normal(0.0, scale))

    # Deterministic: posterior inclusion probabilities P(z_j=1 | beta_j, gamma)
    log_gamma   = jnp.log(gamma)
    log_1m_gamma = jnp.log1p(-gamma)
    log_slab     = dist.Normal(0.0, slab_sd  * omegaCL).log_prob(beta)
    log_spike    = dist.Normal(0.0, spike_sd * omegaCL).log_prob(beta)
    num   = log_gamma + log_slab
    denom = logsumexp(jnp.stack([log_gamma + log_slab,
                                 log_1m_gamma + log_spike]), axis=0)
    incl_prob = jnp.exp(num - denom)
    numpyro.deterministic("incl_prob", incl_prob)
    return beta
\end{lstlisting}

\begin{lstlisting}[language=Python, caption={L1-ball prior for SNP effects in NumPyro}, label={lst:l1ball}]
import numpyro
import numpyro.distributions as dist
import jax.numpy as jnp
from jax import lax

def l1_ball_projection(eta: jnp.ndarray, r: float) -> jnp.ndarray:
    """
    Euclidean projection of eta onto the L1 ball:
      If ||eta||_1 <= r -> beta = eta
      Else beta_i = sign(eta_i) * max(|eta_i| - mu, 0), with mu s.t. sum max(..)=r
    """
    abs_eta = jnp.abs(eta)
    l1 = jnp.sum(abs_eta)

    def _project(_):
        a = jnp.sort(abs_eta)[::-1]
        cumsum = jnp.cumsum(a)
        j = jnp.arange(1, eta.size + 1)
        cond = a > (cumsum - r) / j
        c = jnp.maximum(jnp.sum(cond).astype(jnp.int32), 1)
        mu = (cumsum[c - 1] - r) / c
        return jnp.sign(eta) * jnp.maximum(abs_eta - mu, 0.0)

    return lax.cond(l1 <= r, lambda _: eta, _project, operand=None)

def l1_ball_prior(nSNP: int, omegaCL: float, b_eta: float, r_lambda: float):
    """
    L1-ball prior (noise-scaled):
      b_eta_eff    = b_eta * omegaCL
      r_lambda_eff = r_lambda / omegaCL
      eta_i ~ Laplace(0, b_eta_eff)
      r     ~ Exponential(rate=r_lambda_eff)
      beta  = Proj_{||.||_1 <= r}(eta)
    """
    b_eta_eff = b_eta * omegaCL
    r_lambda_eff = r_lambda / omegaCL
    eta = numpyro.sample("eta", dist.Laplace(0.0, b_eta_eff).expand([nSNP]))
    r   = numpyro.sample("r", dist.Exponential(r_lambda_eff))

    beta = l1_ball_projection(eta, r)
    numpyro.deterministic("beta", beta)
    return beta
\end{lstlisting}
\newpage
\begin{lstlisting}[language=Python, caption={Hierarchical lasso prior for SNP effects in NumPyro}, label={lst:hlasso}]
import numpyro
import numpyro.distributions as dist

def hierarchical_lasso_prior(nSNP, scale_tau, rho, omegaCL):
    # Step 1: auxiliary normals
    z = numpyro.sample("z", dist.Normal(0, 1).expand([nSNP]))

    # Step 2: local shrinkage (Exponential: Laplace-like tails)
    lambda = numpyro.sample("lambda", dist.Exponential(rho).expand([nSNP]))

    # Step 3: global shrinkage (noise-scaled)
    tau_raw = numpyro.sample("tau_raw", dist.HalfNormal(1.0))
    tau = scale_tau * tau_raw * omegaCL

    # Step 4: coefficients beta_j
    beta = z * lambda * tau
    numpyro.deterministic("beta", beta)
    return beta
\end{lstlisting}

\begin{lstlisting}[language=Python, caption={Regularized horseshoe prior for SNP effects in NumPyro. This alternative formulation is based on~\citep{Piironen_2017} (Appendix C.2)}, label={lst:horseshoe}]
import numpyro
import numpyro.distributions as dist
import jax.numpy as jnp

def horseshoe_prior(nSNP, scale_global, nu_global, nu_local,
                    slab_sd, slab_df, omegaCL):
    # Step 1: auxiliary normals
    z = numpyro.sample("z", dist.Normal(0, 1).expand([nSNP]))

    # Step 2: local shrinkage components (half-t via Normal x InvGamma)
    r1_local = numpyro.sample("r1_local", dist.Normal(0, 1).expand([nSNP]))
    r2_local = numpyro.sample("r2_local",
                              dist.InverseGamma(0.5 * nu_local,
                                                0.5 * nu_local)
                              .expand([nSNP]))

    # Step 3: global shrinkage components
    r1_global = numpyro.sample("r1_global", dist.Normal(0, 1))
    r2_global = numpyro.sample("r2_global",
                               dist.InverseGamma(0.5 * nu_global,
                                                 0.5 * nu_global))

    # Step 4: slab component
    caux = numpyro.sample("caux",
                          dist.InverseGamma(0.5 * slab_df,
                                            0.5 * slab_df))

    # Step 5: assemble scales (noise-scaled)
    tau = r1_global * jnp.sqrt(r2_global) * scale_global * omegaCL
    c   = slab_sd * jnp.sqrt(caux) * omegaCL
    lam = r1_local * jnp.sqrt(r2_local)

    # Regularized local scale
    lam_tilde = jnp.sqrt(c**2 * lam**2 / (c**2 + tau**2 * lam**2))

    # Step 6: coefficients beta_j
    beta = z * lam_tilde * tau
    numpyro.deterministic("beta", beta)
    return beta
\end{lstlisting}

\newpage

\begin{lstlisting}[language=Python, caption={R2-D2 prior for SNP effects in NumPyro}, label={lst:r2d2}]
import numpyro
import numpyro.distributions as dist
import jax.numpy as jnp

def _dirichlet_stick_breaking(conc: float, p: int):
    # Exact stick-breaking for symmetric Dirichlet(conc,...,conc)
    k = jnp.arange(p - 1)                      # 0..p-2
    alpha_k    = jnp.full((p - 1,), conc)
    alpha_rest = conc * (p - 1 - k)
    v = numpyro.sample("v", dist.Beta(alpha_k, alpha_rest))  # (p-1,)

    one_minus_v = 1.0 - v
    cumprod = jnp.cumprod(one_minus_v)
    first  = v[0]
    middle = v[1:] * cumprod[:-1]
    last   = cumprod[-1]
    lambda = jnp.concatenate([first[None], middle, last[None]])
    return lambda, v

def r2d2_prior(nSNP: int, a: float, b: float, alpha: float, omega_ref: float):
    # Step 1: global shrinkage
    R2 = numpyro.sample("R2_genetic", dist.Beta(a, b))

    # Step 2: local weights lambda ~ Dirichlet(alpha/p) (stick-breaking)
    conc = alpha / nSNP
    lambda, _ = _dirichlet_stick_breaking(conc, nSNP)

    # Step 3: total prior variance (noise-scaled reference omega_ref)
    tau2 = (R2 / (1.0 - R2)) * (omega_ref ** 2)

    # Step 4: coefficients beta_j
    z = numpyro.sample("z_beta", dist.Normal(0.0, 1.0).expand([nSNP]))
    beta = jnp.sqrt(tau2) * jnp.sqrt(lambda) * z

    numpyro.deterministic("lambda", lambda)
    numpyro.deterministic("beta", beta)
    return beta
\end{lstlisting}

\end{document}